%% file: arxiv_draft.tex
\documentclass[sigplan,screen]{acmart}
\authorsaddresses{}
\renewcommand\footnotetextcopyrightpermission[1]{} 
\setcopyright{acmcopyright}
\acmPrice{15.00}
\acmDOI{10.1145/3192366.3192412}
\acmYear{2018}
\copyrightyear{2018}
\acmISBN{978-1-4503-5698-5/18/06}
\acmConference[PLDI'18]{39th ACM SIGPLAN Conference on Programming Language Design and Implementation}{June 18--22, 2018}{Philadelphia, PA, USA}

\startPage{1}

\setcopyright{none}

\usepackage{booktabs}
\usepackage{subcaption}

\usepackage{url}
\usepackage{paralist}
\usepackage[small,compact]{titlesec}
\usepackage{setspace}
\usepackage{times}
\usepackage{amsmath,amssymb,amsfonts,amsthm}
\usepackage[T1]{fontenc}
\usepackage[latin9]{inputenc}
\usepackage{color}
\usepackage{listings}
\usepackage{float}
\usepackage{graphics}
\usepackage{graphicx}
\usepackage{wrapfig}
\usepackage{pifont}
\usepackage{epsfig}
\usepackage[linesnumbered,ruled,vlined]{algorithm2e}
\usepackage{rotating}
\usepackage{mathpartir}
\usepackage{array}
\usepackage[percent]{overpic}
\usepackage{ragged2e}
\usepackage{fancybox}
\usepackage{multirow}
\usepackage{tikz, pgfplots}
\usetikzlibrary{arrows.meta}

\usepackage{subcaption}
\usepackage{placeins}
\usepackage{float}
\usepackage{booktabs}
\usepackage[final]{changes} 
\usepackage{flushend}
\definechangesauthor[name={Meital Zilberstein}, color=magenta]{mz}
\definechangesauthor[name={Eran Yahav}, color=blue]{ey}
\definechangesauthor[name={Uri Alon}, color=green]{ua}

\input{genericmacros}
\input{specificmacros}

\renewcommand{\textrightarrow}{$\rightarrow$}

\begin{document}


\defcitealias{JavaParser}{JavaParser}
\defcitealias{UnuglifyJS}{UnuglifyJS}
\defcitealias{UglifyJS}{UglifyJS}
\defcitealias{JSNice}{JSNice}
\defcitealias{Roslyn}{Roslyn}




\title{A General Path-Based Representation\\for Predicting Program Properties}


\author{Uri Alon}
\affiliation{
  \institution{Technion}            
  \city{Haifa}
  \country{Israel}                    
}
\email{urialon@cs.technion.ac.il}          

\author{Meital Zilberstein}
\affiliation{
  \institution{Technion}            
  \city{Haifa}
  \country{Israel}                    
}
\email{mbs@cs.technion.ac.il}          

\author{Omer Levy}
\affiliation{
  \institution{University of Washington}            
  \city{Seattle}
  \state{WA}
}
\email{omerlevy@cs.washington.edu}          

\author{Eran Yahav}
\affiliation{
  \institution{Technion}            
  \city{Haifa}
  \country{Israel}                    
}
\email{yahave@cs.technion.ac.il}          


\input{abstract3}



\maketitle

\input{intro2}

\input{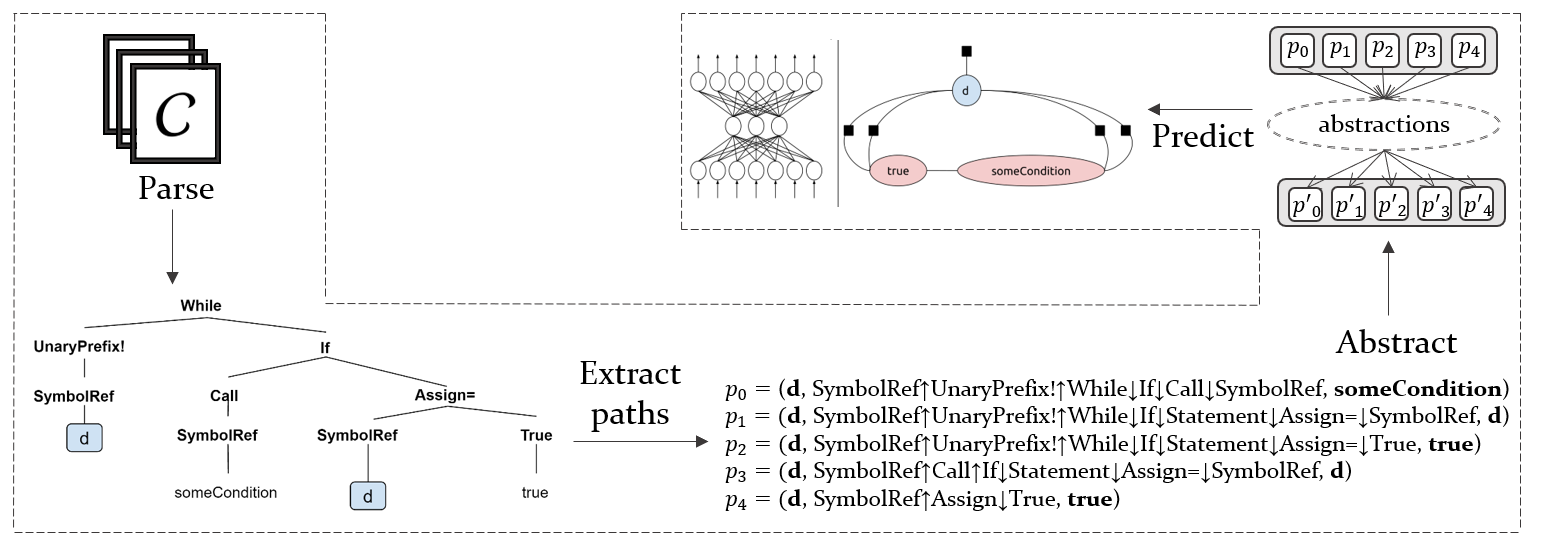}

\input{background}

\input{application}
\input{evaluation}
\input{relatedWork}
\input{conclusion}
\input{acknowledgement}

\begingroup
\let\clearpage\relax
\input{arxiv_draft.bbl}

\endgroup





\end{document}

%% file: genericmacros.tex
\usepackage{ifthen}
\newcommand{\todoi}[1]{}

\newcommand{\COMMENT}[1]{}
\newcommand{\ignore}[1]{}

\newcommand{\figref}[1]{Fig.~\ref{Fi:#1}}

\newcommand{\tabref}[1]{Table~\ref{Ta:#1}}
\newcommand{\secref}[1]{Section~\ref{Sec:#1}}

\newcommand{\ssecref}[1]{Sec.~\ref{Sec:#1}}

\newcommand{\figlabel}[1]{\label{Fi:#1}}

\newcommand{\tablabel}[1]{\label{Ta:#1}}
\newcommand{\seclabel}[1]{\label{Sec:#1}}

\newcommand{\equlabel}[1]{\label{Eq:#1}}

\theoremstyle{definition}

\newboolean{TR}
\setboolean{TR}{false}
\ifthenelse{\boolean{TR}}{

\newcommand{\TrOnly}[1]{#1}
\newcommand{\SubOnly}[1]{}
\newcommand{\TrOnlyInFootnote}[1]{#1}
\newcommand{\TrOnlyInTable}[1]{#1}}
{

\newcommand{\TrOnly}[1]{}
\newcommand{\SubOnly}[1]{#1}
\newcommand{\TrOnlyInFootnote}[1]{}
\newcommand{\TrOnlyInTable}[1]{}}



%% file: specificmacros.tex
\definecolor{airforceblue}{rgb}{0.36, 0.54, 0.66}
\definecolor{awesome}{rgb}{1.0, 0.13, 0.32}
\definecolor{blue(pigment)}{rgb}{0.2, 0.2, 0.6}
\definecolor{britishracinggreen}{rgb}{0.0, 0.26, 0.15}
\definecolor{cadmiumgreen}{rgb}{0.0, 0.42, 0.24}
\definecolor{carminered}{rgb}{1.0, 0.0, 0.22}
\definecolor{electricindigo}{rgb}{0.44, 0.0, 1.0}
\definecolor{mygray}{rgb}{0.5,0.5,0.5}
\hypersetup{colorlinks=true,linkcolor=blue(pigment), citecolor=cadmiumgreen, urlcolor=electricindigo}

\lstdefinestyle{nonumbering}{
        basicstyle=\ttfamily\scriptsize,
        language=C,
        emph={},
        emphstyle={\underbar},
        numbers=none,
        numberfirstline=false,
        numberstyle=\tiny,
        escapeinside={(*@}{@*)}
}
\lstdefinestyle{small}{
        basicstyle=\tiny,
        language=C,
        emph={},
        emphstyle={\underbar},
        numbers=none,
        numberfirstline=false,
        numberstyle=\tiny,
        escapeinside={(*@}{@*)}
}
\lstdefinelanguage
   [x64]{Assembler}     
   [x86masm]{Assembler} 
   {morekeywords={CDQE,CQO,CMPSQ,CMPXCHG16B,JRCXZ,LODSQ,MOVSXD, %
                  POPFQ,PUSHFQ,SCASQ,STOSQ,IRETQ,RDTSCP,SWAPGS, %
                  rax,rdx,rcx,rbx,rsi,rdi,rsp,rbp, %
                  MEM, FLAGS, FLG, assert, %
                  r9,r9d,r12,r12d,r12b, r13,r13d,r14,r14d,r15,r15d}} 
\lstdefinestyle{asm}{
        basicstyle=\ttfamily\scriptsize,
        language=[x64]Assembler,
        emph={},
        emphstyle={\underbar},
        numbers=none,
        numberfirstline=false,
        numberstyle=\tiny,
        escapeinside={(*@}{@*)}
}
\lstdefinestyle{asm-small}{
        basicstyle=\ttfamily\tiny,
        language=[x64]Assembler,
        emph={},
        emphstyle={\underbar},
        numbers=none,
        numberfirstline=false,
        numberstyle=\tiny,
        escapeinside={(*@}{@*)}
}
\lstdefinestyle{asm-numbered}{
        basicstyle=\ttfamily\scriptsize,
        language=[x64]Assembler,
        emph={},
        emphstyle={\underbar},
        numbers=left,
        numberfirstline=false,
        numberstyle=\tiny\color{mygray},
        escapeinside={(*@}{@*)}
}

\usepackage{xspace} 

\newcommand{\hiddentext}[1]{}

\newcommand{\para}[1]{\vspace{3pt}\noindent\textbf{\textit{#1}}}
\newcommand{\tab}{\hspace*{2em}}

\newcommand{\scode}[1]{{\small \texttt{#1}}}

\renewcommand{\phi}{\varphi}

\lstdefinelanguage{JavaScript}{
  keywords={break, case, catch, continue, debugger, default, delete, do, else, false, finally, for, function, if, in, instanceof, new, null, return, switch, this, throw, true, try, typeof, var, void, while, with},
  morecomment=[l]{//},
  morecomment=[s]{/*}{*/},
  morestring=[b]',
  morestring=[b]",
  ndkeywords={class, export, boolean, throw, implements, import, this},
  keywordstyle=\color{blue},
  ndkeywordstyle=\color{darkgray}\bfseries,
  identifierstyle=\color{black},
  commentstyle=\color{purple}\ttfamily,
  stringstyle=\color{red}\ttfamily,
  sensitive=true
}

\newcommand{\tool}{\textsc{Pigeon}\xspace}
\newcommand{\maxlen}{{$max\_length$}\xspace}
\newcommand{\maxwidth}{{$max\_width$}\xspace}

\newcommand{\javaacc}{58.2\%}
\newcommand{\jsacc}{67.3\%}
\newcommand{\pythonacc}{56.7\%} 

%% file: abstract3.tex
\begin{abstract}
Predicting program properties such as names or expression types has a wide range of applications. It can ease the task of programming, and increase programmer productivity. A major challenge when learning from programs is \emph{how to represent programs in a way that facilitates effective learning}.

We present a \emph{general path-based representation} for learning from programs. Our representation is purely syntactic and extracted automatically. The main idea is to represent a program using paths in its abstract syntax tree (AST). This allows a learning model to leverage the structured nature of code rather than treating it as a flat sequence of tokens.

We show that this representation is general and can:
\begin{inparaenum}[(i)]
\item cover different prediction tasks,
\item drive different learning algorithms (for both generative and discriminative models), and
\item work across different programming languages.
\end{inparaenum}

We evaluate our approach on the tasks of predicting variable names, method names, and full types. We use our representation to drive both CRF-based and word2vec-based learning, for programs of four languages: JavaScript, Java, Python and C\#. Our evaluation shows that our approach obtains better results than task-specific handcrafted representations across different tasks and programming languages.
\end{abstract} 	

%% file: intro2.tex
\section{Introduction}\seclabel{Intro}

Leveraging machine learning models for predicting program properties such as variable names, method names, and expression types is a topic of much recent interest~\cite{jsnice2015,allamanis2015,conv16, decisionTrees2016, phog16, Maddison:2014:SGM:3044805.3044965}. These techniques are based on learning a statistical model from a large amount of code and using the model to make predictions in new programs. A major challenge in these techniques (and in many other machine-learning problems) is how to represent instances of the input space to facilitate learning~\cite{shwartz2014}. Designing a program representation that enables effective learning is a critical task that is \emph{often done manually for each task and programming language}.

\para{Our approach} We present a novel program representation for learning from programs. Our approach uses different path-based abstractions of the program's abstract syntax tree. This family of path-based representations is natural, general, fully automatic, and works well across different tasks and programming languages.

\para{AST paths} We define AST paths as paths between nodes in a program's abstract syntax tree (AST). To automatically generate paths, we first parse the program to produce an AST, and then extract paths between nodes in the tree. We represent a path in the AST as a sequence of nodes connected by up and down movements, and represent a program element as the set of paths that its occurrences participate in. \figref{code-example} shows an example JavaScript program. \figref{programPath} shows its AST, and one of the extracted paths. The path from the first occurrence of the variable \scode{d} to its second occurrence can be represented as:
{\small
\[
\textrm{SymbolRef }{\uparrow}\textrm{ UnaryPrefix! }{\uparrow}\textrm{ While }{\downarrow}\textrm{ If }{\downarrow}\textrm{ Assign= }{\downarrow}\textrm{ SymbolRef}
\]
}
This is an example of a pairwise path between leaves in the AST, but in general the family of path-based representations contains n-wise paths, which do not necessarily span between leaves and do not necessarily contain all the nodes in between. Specifically, we consider several choices of subsets of this family in~\secref{application}.

\input{intro-ast-path}

Using a path-based representation has several major advantages over existing methods:
\begin{enumerate}
\item Paths are generated automatically: there is no need for manual design of features aiming to capture potentially interesting relationships between program elements. This approach extracts unexpectedly useful paths, without the need for an expert to design features. The user is required only to choose a subset of our proposed family of path-based representations.
\item This representation is useful for any programming language, without the need to identify common patterns and nuances in each language.
\item The same representation is useful for a variety of prediction tasks, by using it with off-the-shelf learning algorithms or by simply replacing the representation of program elements in existing models (as we show in~\secref{results}).
\item AST paths are purely syntactic, and do not require any semantic analysis.
\end{enumerate}


\paragraph{Tasks} In this paper, we demonstrate the power and generality of AST paths on the following tasks:
\begin{itemize}
\item \textbf{Predicting names for program elements} Descriptive and informative names for program elements such as variables and classes play a significant role in the readability and comprehensibility of code. Empirical studies have shown that choosing appropriate names makes code more understandable~\cite{takang96}, reduces code maintenance efforts, and leads to fewer bugs~\citep{butler2009}. A study in the Psychology of Programming suggests that the ways in which programmers choose names reflect deep cognitive and linguistic influences~\cite{liblit2006}. A meaningful name describes the role of a program element, carries its semantic meanings, and indicates its usage and behavior throughout the program.
\item \textbf{Predicting method names} Good method names adequately balance the need to describe the internal implementation of the method and its external usage~\cite{Host:2009:DMN:1615184.1615204}. When published in a popular library's API, descriptive and intuitive method names facilitate the use of methods and classes, while poorly chosen names can doom a project to irrelevance~\cite{allamanis2015}. Although method names are clearly program elements and can be predicted by the previous task, in this task we assumes that all the other names in the method are given, along with the names of the elements around the method invocation, when available in the same file.
\item \textbf{Predicting expression types} Statistical type prediction allows (likely) types of expressions to be inferred without the need for type inference, which often requires a global program view (possibly unavailable, e.g., in the case of snippets from sites such as StackOverflow). 
\end{itemize}


\begin{figure*}
\centering
\includegraphics[width=7in]{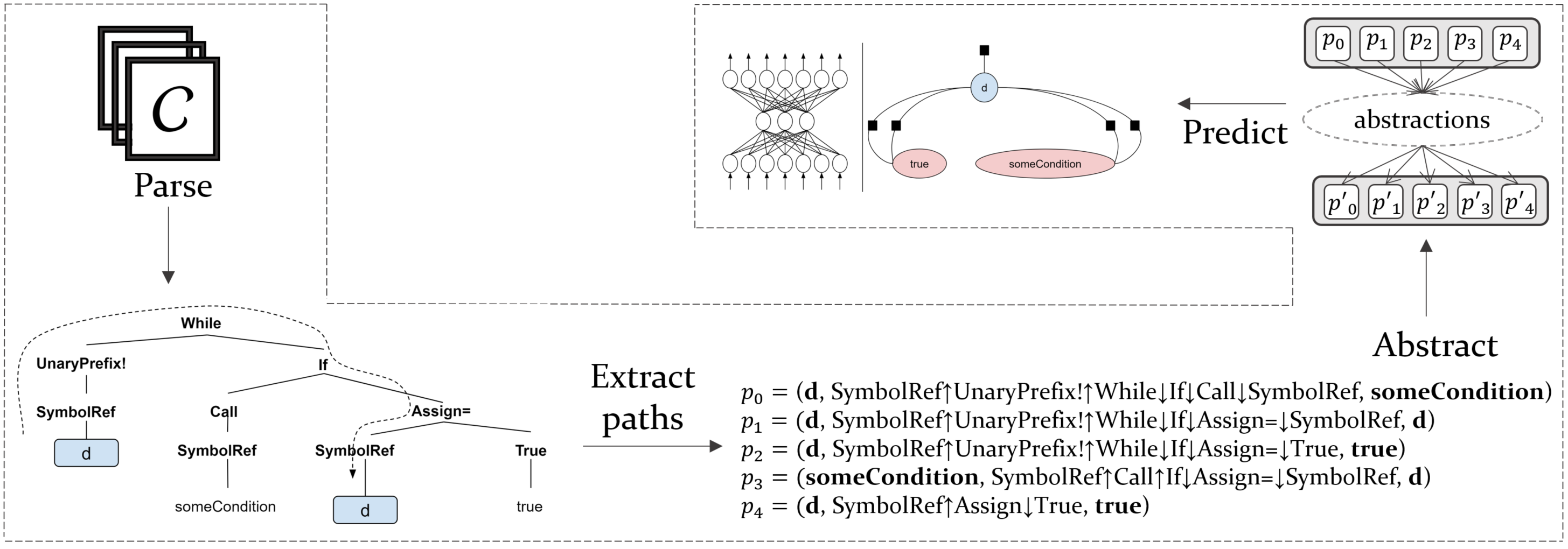}
\caption{An overview of our approach. We start with a code snippet $\mathcal{C}$, and extract its path representation to be used as an input to machine learning models. The AST and paths were extracted from the example program in~\figref{code-example}.}
\label{Fi:overview_fig}
\end{figure*}

\citet{jsnice2015} used relations in the AST as features for learning tasks over programs. They defined an explicit grammar to derive features which capture specific relationships between nodes in the AST of JavaScript programs, as well as relations produced by language-specific semantic analysis, such as ``may call'' and ``may access''. We show that our \emph{automatic general representation} performs better than their features for their original task, and also generalizes to drive two different learning algorithms and three different prediction tasks, over different programming languages.

Paths in an AST have also been used by~\citet{phog16} and by~\citet{decisionTrees2016,raychev2016noisy} for a different goal: identifying context nodes. These works do not use the paths themselves as a representation of the input, and the prediction is only affected by the context node that was found on the other end of the path.  In our work, we use the path itself as a representation of a program element. Therefore, the prediction depends not only on the context node but also on \emph{the way it is related} to the element in question.

\citet{allamanis2015} defined the challenging task of predicting method names, which can be viewed as a form of function summarization~\cite{conv16}. We show that our representation performs better by being able to learn across projects.
	
We provide a more elaborate discussion of related work, including deep learning approaches, in~\secref{relatedWork}.

\para{Contributions} The main contributions of this paper are:
\begin{itemize}
\item A new, general family of representations for program elements. The main idea is to use AST paths as representations of code.
\item A cross-language tool called \tool, which is an implementation of our approach for predicting program element names, method names, and types.
\item An evaluation on real-world programs. Our experiments show that our approach produces accurate results for different languages (JavaScript, Java, Python, C\#), tasks (predicting variable names, method names, types) and learning algorithms (CRFs, word2vec). Furthermore, for JavaScript and Java, where previous methods exist, our automatic approach produces more accurate results.
\end{itemize}

%% file: intro-ast-path.tex
\begin{figure}
\hspace{10mm}
\begin{subfigure}[b]{0.42\textwidth}
\lstset{language=JavaScript, basicstyle=\footnotesize\ttfamily,emph={fence},emphstyle=\underbar,escapeinside={(*@}{@*)}}
\begin{lstlisting}
while (!d) {
	if (someCondition()) {
		d = true;
	}
}
\end{lstlisting}
\caption{A simple JavaScript program.}
\label{Fi:code-example}
\end{subfigure}
\\
\begin{subfigure}[b]{0.42\textwidth}
\includegraphics[width=3in]{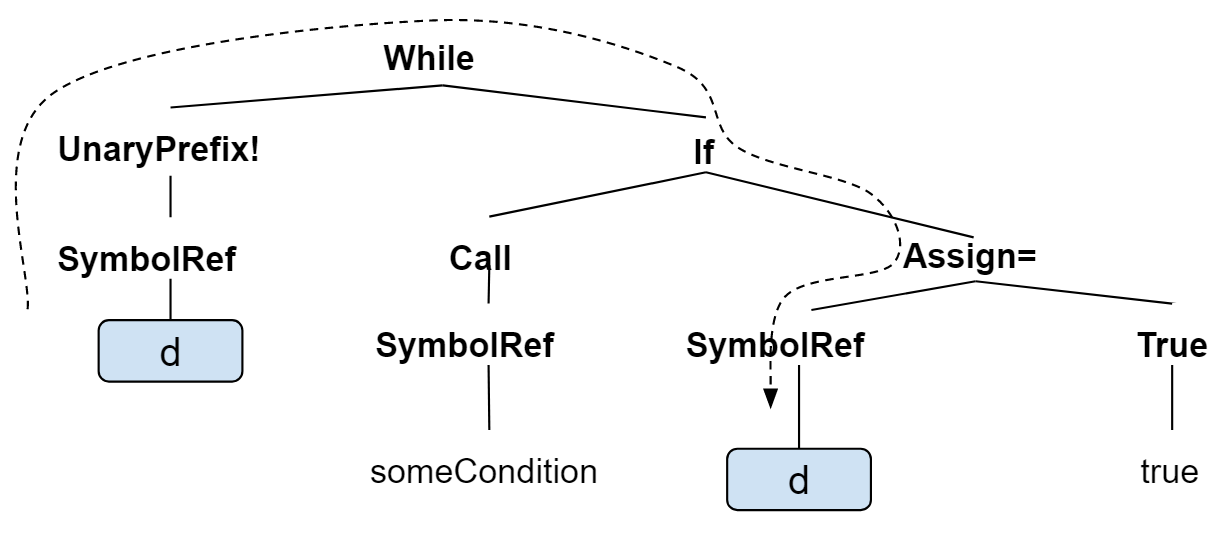}
\caption{The program's AST, and example to an AST path.}
\label{Fi:programPath}
\end{subfigure}
\caption{A JavaScript program and its AST, along with an example of one of the paths.}
\end{figure} 

%% file: overview.tex
\section{Overview}\label{Se:Overview}

In this section, we illustrate our approach with a simple JavaScript program for the task of predicting names; as we show in later sections, the same approach also applies to other tasks, other languages, and other learning algorithms.

Given a program with non-descriptive names, our goal is to predict likely names for local variables and function parameters. The non-descriptive names could have been given by an inexperienced programmer, or could have been the result of deliberate stripping. In the latter case, we refer to such a program as a program with \emph{stripped names}. Stripping names can be part of a minification process in JavaScript, or obfuscation in Java and other languages.

Consider the code snippet of~\figref{code-example}. This simple snippet captures a common programming pattern in many languages. Suppose that we wish to find a better name for the variable \scode{d}.

\para{Program element representation} The main idea of our approach is to extract paths from the program's AST and use them to represent an element, such as the variable \scode{d}, in a machine learning model. \figref{overview_fig} shows an overview of this process.
First, we parse the query program to construct an AST. Then, we traverse the tree and extract paths between AST nodes. To simplify presentation, in this example we only consider pairwise paths between AST leaves. We assume that a path is represented as a sequence of AST nodes, linked by up and down movements (denoted by arrows). As we describe in~\ssecref{application}, the path can also connect a leaf and a higher nonterminal in the AST, connect several nodes (n-wise path), and can be abstracted in different levels.

Consider the $p_{1}$ in~\figref{overview_fig}, between the two  occurrences of the variable \scode{d}:
\begin{equation}
\footnotesize
\textrm{SymbolRef }{\uparrow}\textrm{ UnaryPrefix! }{\uparrow}\textrm{ While }{\downarrow}\textrm{ If }{\downarrow}\textrm{ Assign= }{\downarrow}\textrm{ SymbolRef}
\tag{I}\label{eq:dTodPath}
\end{equation}
The path expresses the fact that the variable \scode{d} is used, with negation, as a stopping condition of a ``while'' loop, and then assigned a new value if an ``if'' condition inside the loop evaluates to \scode{true}. This path alone expresses the fact that \scode{d} is the stopping condition of the loop. 

The path $p_{4}$ in~\figref{overview_fig}, between the variable \scode{d} and the value \scode{true} is:
\begin{equation}
\footnotesize
\textrm{SymbolRef }{\uparrow}\textrm{ Assign= }{\downarrow}\textrm{True}
\tag{II}\label{eq:dToTruePath}
\end{equation}
This path captures the fact that the assignment changes the value of \scode{d} to \scode{true}, and therefore it is indeed the assignment that stops the loop.

\para{Prediction}
By observing these two paths, a programmer is likely to name \scode{d} ``done'', ``complete'', ``stop'' or something similar. Indeed, a learning model that was trained using our representation predicts that the most likely name for the variable is \scode{done}, and neither ``done'', ``complete'', nor any similar name was predicted by past work 
for this example. 

\para{Learning algorithms}
The learning model can vary between different algorithms, presenting tradeoffs of efficiency and accuracy. In~\secref{results} we show that both CRFs and word2vec can be used for this prediction task.
In both of these learning algorithms, using AST paths produces better results than the alternative representations, whether they are manually designed or sequence-based representations.

\para{Path abstractions} Automatic generation may produce a prohibitively large number of paths. To control the number of paths, \emph{higher levels of abstraction} can be used. Instead of representing the whole path node-by-node, it can be further abstracted by keeping only parts of it, which results in similar paths being represented equally, as we show in~\secref{Abstractions}. Another way to control the number of paths is to limit the number of extracted paths. We provide hyper-parameters (i.e., model configurations that are not tuned by the optimization process) that control the maximal length and width of AST paths.
The number of extracted paths can be further reduced using downsampling, with minimal impact on accuracy and a significant saving in training time (\ssecref{results}). These methods make the accuracy -- training time tradeoff tunable.

\begin{figure}[]
\begin{subfigure}[t]{0.25\textwidth}
\lstset{language=JavaScript, basicstyle=\footnotesize\ttfamily,emphstyle=\underbar,escapeinside={(*}{*)}}
\begin{lstlisting}
var (*\bfseries d*) = false;
while(!(*\bfseries d*)) {
    doSomething();
    if (someCondition()) {
        (*\bfseries d*) = true;
    }
}
\end{lstlisting}
\caption{}
\figlabel{fullDone}
\end{subfigure}%
\unskip \vrule
\hspace{3mm}
\begin{subfigure}[t]{0.2\textwidth}
\lstset{language=JavaScript, basicstyle=\footnotesize\ttfamily,emphstyle=\underbar,showlines=true,escapeinside={(*}{*)}}
\begin{lstlisting}

someCondition();
doSomething();
var (*\bfseries d*) = false;
(*\bfseries d*) = true;

\end{lstlisting}
\caption{}
\figlabel{indistinguishableDone}
\end{subfigure}
\caption{An example for two code snippets that are indistinguishable by the model of~\citet{jsnice2015}, and are easily distinguishable by AST paths.}
\figlabel{similarWhile}
\end{figure}

\para{The discriminative power of AST paths} Examples indistinguishable by manually designed representations will always exist. For example, \citetalias{UnuglifyJS}~\cite{jsnice2015} extracts an identical set of relations (and therefore predicts the same name for \scode{d}) for the example in~\figref{fullDone} and for the simplified example in~\figref{indistinguishableDone}, even though the variable \scode{d} clearly does not play a similar role in these two examples. In contrast, these two examples are easily distinguishable using our AST paths.


\para{Key aspects} The example highlights several key aspects of our approach:
\begin{itemize}
\item Useful paths such as path~\ref{eq:dTodPath} span multiple lines of the program, but are also supported by shorter paths like path~\ref{eq:dToTruePath}, which only spans a single program line. Short paths alone are not enough to predict a meaningful name. Making a prediction using all paths that an element participates in provides a rich context for predicting the name of the element.
\item No special assumptions regarding the AST or the programming language were made, making the same mechanism useful in other languages in a similar way. 
\item This representation can be plugged into existing models as a richer representation of the input code, without interfering with the learning algorithm itself. 
\item AST paths can distinguish between programs that previous works could not.
\item In addition to predicting \scode{done}, a model trained with AST paths can propose several semantically similar names, as we demonstrate in~\ssecref{results}. This shows that AST paths are strong indicators of the program element's semantics.
\end{itemize}

%% file: background.tex
\section{Background}\label{Se:background}
In this section, we provide necessary background.
In Sections~\ref{Sec:crfs} and~\ref{Sec:w2v} we present CRFs and word2vec and how they are used to predict program properties.

\subsection{Conditional Random Fields}\seclabel{crfs}
Probabilistic graphical models are a formalism for expressing the dependence structure of entities. Traditionally, graphical models have been used to represent the joint probability distribution $P\left(y,x\right)$, where \emph{y} represents an assignment of attributes for the entities that we wish to predict, and \emph{x} represents our observed knowledge about the entities~\cite{sutton2012, pearl2011bayesian, pearl2014probabilistic}. Such models are called \emph{generative} models. However, modeling the joint distribution requires modeling the marginal probability $P\left(x\right)$, which can be difficult, computationally expensive, and in our case requires us to estimate the distribution of programs~\cite{jsnice2015}.

A simpler solution is to model the conditional distribution $P\left(y|x\right)$ directly. Such models are called \emph{discriminative} models. This is the approach taken by \emph{Conditional Random Fields} (CRFs). A CRF is a conditional distribution $P\left(y|x\right)$ with an associated graphical structure~\cite{lafferty2001}. They have been used in several fields such as natural language processing, bioinformatics, and computer vision.

Formally, given a variable set $Y$ and a collection of subsets $\left\{ Y_{a}\right\} _{a=1}^{A}$ of $Y$, an \emph{undirected graphical model} is the set of all distributions that can be written as:
\begin{equation}
P\left(y\right)=\frac{1}{Z}\prod_{a=1}^{A}\Psi_{a}\left(y_{a}\right)
\end{equation}
where each $\Psi_{a}\left(y_{a}\right)$ represents a \emph{factor}. Each factor $\Psi_{a}\left(y_{a}\right)$ depends only on a subset $Y_{a}\subseteq Y$ of the variables. Its value is a non-negative scalar which can be thought of as a measure of how compatible the values $y_{a}$ are. The constant $Z$ is a normalization factor, also known as a partition function, that ensures that the distribution sums to $1$. It is defined as:
\begin{equation}
Z=\sum_{y}\prod_{a=1}^{A}\Psi_{a}\left(y_{a}\right)
\end{equation}

A CRF can also be represented as a bipartite undirected graph $G=\left(V,F,E\right)$, in which one set of nodes $V=\left\{ 1,2,...,\left|Y\right|\right\}$ represents indices of random variables, and the other set of nodes $F=\left\{ 1,2,...,A\right\}$  represents indices of the factors.

Several algorithms and heuristics were suggested for training CRFs and finding the assignment that yields the maximal probability~\cite{lafferty2001, koller2007, sutton2012}. We will not focus on them here, since our work is orthogonal to the learning model and the prediction algorithm.

Using CRFs to predict program properties was first proposed by~\citet{jsnice2015}, where each node represented an identifier in the code. These include variables, constants, function and property names. The factors represented the relationships or dependencies between those identifiers, and were \emph{defined by an explicit grammar and relations that were produced using semantic analysis}.

To demonstrate the use of AST paths with CRFs, we use CRFs exactly as they were used by~\citet{jsnice2015} but use AST paths instead of their original factors. Additionally, we introduce the use of unary factors (factors that depend on a single node). Unary factors are derived automatically by AST paths between different occurrences of the same program element throughout the program.

\subsection{Neural Word Embeddings}\seclabel{w2v}

Recently, neural-network based approaches have shown that syntactic and semantic properties of natural language words can be captured using low-dimensional vector representations, referred to as ``word embeddings'', such that similar words are assigned similar vectors~\cite{bengio2003,collobert2008,glove2014}. These representations are trained over large swaths of unlabeled text, such as Wikipedia or newswire corpora, and are essentially unsupervised. The produced vectors were shown to be effective for various NLP tasks~\cite{turian2010,collobert2011}.

In particular, the skip-gram model trained with the negative sampling objective (SGNS), introduced by~\citet{mikolovEfficient2013,mikolovDistributed2013}, has gained immense popularity via the word2vec toolkit, and substantially outperformed previous models while being efficient to train.

SGNS works by first extracting the \emph{contexts}: $c_1, ..., c_n$ of each word $w$. It then learns a latent $d$-dimensional representation for each word and context in the vocabulary ($\vec{w}, \vec{c} \in \mathbb{R}^d$) by maximizing the similarity of each word $w$ and context $c$ that were observed together, while minimizing the similarity of $w$ with a randomly sampled context $c'$. In Mikolov et al.'s original implementation, each context $c_i$ is a neighbor of $w$, i.e., a word that appeared within a fixed-length window of tokens from $w$. \citet{levy2014dependency} extended this definition to include arbitrary types of contexts.

As shown by~\citet{levy2014neural}, this algorithm implicitly tries to encode the pointwise mutual information (PMI) between each word and context via their vector representations' inner products:
\begin{equation}
\vec{w}\cdot\vec{c} = PMI(w,c) = \log\frac{p(w,c)}{p(w)p(c)}
\label{Eq:skipgram}
\end{equation}
where each probability term models the frequency of observing a word $w$ and a context $c$ together (or independently) in the training corpus.


Recently, a simple model has achieved near state-of-the-art results for the lexical substitution task using embeddings that were learned by word2vec~\cite{melamud15}. The task requires identifying meaning-preserving substitutes for a target word in a given sentential context. The model in this work uses both word embeddings and context embeddings, and looks for a word out of the entire vocabulary whose embedding is the closest to all the given context embeddings and to the original word.
The similarities between the substitute word and each of the contexts and the original word are aggregated by an arithmetic mean or a geometric mean.

In contrast to natural language methods, our method does not use the original word but finds the unknown name by aggregating only the similarities between the candidate vector $w$ and each of the given context vectors $\widetilde{C}$:
\begin{equation}
prediction=argmax_{w\in W}\sum_{c\in \widetilde{C}}\left(w\cdot c\right)
\end{equation}

To demonstrate the use of AST paths with word2vec, we use AST paths as the context of prediction. As we show in~\secref{results}, using AST paths as context gives a relative improvement of $96\%$ over treating code as a token-stream and using the surrounding tokens as context.

%% file: application.tex
\section{AST Paths Representation}\seclabel{application}
In this section, we formally describe the family of AST paths.

\subsection{AST Paths}\seclabel{PathFeatures}
To learn from programs, we are looking for a representation that captures interesting properties of ASTs while keeping it open for generalization. One way to obtain such a representation is to decompose the AST to parts that repeat across programs but can also discriminate between different programs. One such decomposition is into paths between nodes in the AST. We note that in general we consider n-wise paths, i.e., those that have more than two ends, but for simplicity we base the following definitions on pairwise paths between AST terminals.

We start by defining an AST, an AST-path, a path-context and an abstract path-context.

\begin{definition}[Abstract Syntax Tree]
An Abstract Syntax Tree (AST) for a code snippet $\mathcal{C}$ is a tuple $\langle N, T, X, s, \delta, val \rangle$ where $N$ is a set of nonterminal nodes, $T$ is a set of terminal nodes, $X$ is a set of terminal values, $s\in N$ is the root node, $\delta: N\rightarrow \left(N\cup T\right)^*$ is a function that maps a nonterminal node to a list of its children, and $val:T\rightarrow X$ is a function that maps a terminal node to an associated value. Every node except the root appears exactly once in all the lists of children.
\end{definition}

For convenience, we also define $\pi:\left(N\cup T\right)\rightarrow N$, the inverse function for $\delta$. Given a node, this function returns its parent node, such that for every two terminal or nonterminal nodes $y_1,y_2\in\left(N\cup T\right)$, one is the parent node of the other if and only if the latter is in the list of children of the former: $\pi\left(y_1\right)=y_2\iff y_1\in \delta\left(y_{2}\right)$.
In the case of the start symbol, its parent is undefined.

Next, we define AST pairwise paths. For convenience, in the rest of this section we assume that all definitions refer to a single AST $ \langle N, T, X, s, \delta, val \rangle$.

An AST pairwise path is a path between two nodes in the AST, formally defined as follows:

\begin{definition}[AST path]
An AST-path of length $k$ is a sequence $n_{1}d_{1}...n_{k}d_{k}n_{k+1}$, where for $i\in \left[1..k+1\right]$: $n_{i}\in \left(N\cup T\right)$ are terminals or nonterminals and for $i\in \left[1..k\right]$:  $d_{i} \in \{ \uparrow, \downarrow\}$ are movement directions (either up or down in the tree). If $d_{i}=\uparrow$, then: $n_{i+1}=\pi\left(n_i\right)$; if $d_{i}=\downarrow$, then: $n_{i}=\pi\left(n_{i+1}\right)$. We use $start\left(p\right)$ to denote $n_{1}$ and $end\left(p\right)$ to denote $n_{k+1}$.
\end{definition}

We define a \emph{path-context} as a tuple of an AST path and the values associated with its end nodes: (i.e. $n_{1}$ and $n_{k+1}$). In general, we consider path-contexts which span between arbitrary AST nodes, e.g., a terminal and its ancestor, but for simplicity, we base the following definitions on path-contexts which span between terminals:

\begin{definition}[Path-context]
Given an AST Path $p$, its path-context is the triplet $\langle x_{s},p,x_{f} \rangle$ where $x_{s} = val\left(start\left(p\right)\right)$ and $x_{f} = val\left(end\left(p\right)\right)$ are the values associated with the start and end nodes of $p$.
\end{definition}

That is, a path-context describes two nodes from the AST with the syntactic path between them.

Finally, we define an \emph{Abstract path-context} as an abstraction of concrete path context:

\begin{definition}[Abstract path-context]
Given a path-context $\langle x_{s},p,x_{f} \rangle$ and an abstraction function $\alpha:P\rightarrow\hat{P}$,
an abstract path-context is the triplet $\langle x_{s},\alpha\left(p\right),x_{f} \rangle$,
where $P$ is the set of AST paths,
$\hat{P}$ is the set of abstract AST paths,
and $\alpha$ is a function that maps a path to an abstract representation of it.
\end{definition}

The abstraction function $\alpha$ is any function that transforms a path to a different representation. A trivial abstraction function is $\alpha_{id}$, which maps a path to itself: $\alpha_{id}\left(p\right)=p$.

\input{statementpath-figure}

\begin{example}
For example, consider the JavaScript line of code in~\figref{childIdExampleLine} and its partial AST in~\figref{statementPath}. We denote the path between the variable \scode{item} to the variable \scode{array} by $p$. Using $\alpha_{id}$, the abstract path-context of $p$ is:
\begin{align}
\small
&\langle \scode{item},\alpha_{id}\left(p\right),\scode{array} \rangle= \\
&\langle\scode{item},\left(SymbolVar\uparrow VarDef \downarrow Sub\downarrow SymbolRef\right),\scode{array}\rangle
\end{align}
Using a different abstraction function yields a different abstract path-context, for example $\alpha_{forget-arrows}$:
\begin{align}
\small
\langle & \scode{item},\alpha_{forget-arrows}\left(p\right),\scode{array} \rangle= \\
&\langle \scode{item}, \left(SymbolVar , VarDef , Sub , SymbolRef\right), \scode{array} \rangle
\end{align}
\end{example}

Na\"ively extracting all the paths in the AST and representing each of them uniquely can be computationally infeasible, and as a result of the bias-variance tradeoff~\cite[p.~37 and 219]{statisticalLearningBook}, can lead to worse prediction results. However, alternative abstraction functions can be used to control the number of distinct extracted paths. In~\secref{Abstractions} we describe alternative abstractions that abstract some of the information, and thus allow us to tune the trade-off between accuracy, training time, and model size.

\begin{figure}
\begin{subfigure}[b]{0.15\textwidth}
\centering
\lstset{language=JavaScript, basicstyle=\footnotesize\ttfamily,emph={fence},emphstyle=\underbar,showlines=true,escapeinside={(*@}{@*)}}
\begin{lstlisting}

var a, b, c, d;

\end{lstlisting}
\end{subfigure}
\unskip \vrule
\hspace{1mm}
\begin{subfigure}[b]{0.3\textwidth}
\centering
\includegraphics[width=2in]{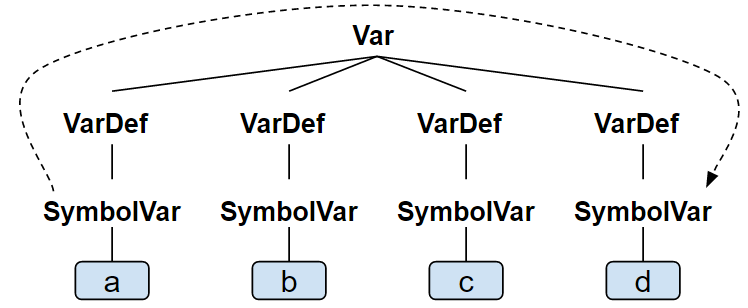}
\end{subfigure}
\caption{An example statement and its AST, with an example of a path between the \emph{SymbolVar} terminals that represent \scode{a} and \scode{d}. The length of this path is $4$, and its width is $3$.}
\label{Fi:length-width}
\end{figure}

\subsection{Limiting the Number of Paths}\seclabel{ExtractingPaths}
Another approach for controlling the number of distinct paths is to limit the \emph{number of extracted paths}.

\para{Path length and width}
We define hyper-parameters that limit the path length and width. We define the following hyper-parameters:
\begin{itemize}
\item \maxlen, defined as the maximal \emph{length} of a path, i.e., the maximum value of $k$.
\item \maxwidth, defined as the maximal allowed difference between sibling nodes that participate in the same path, as shown in~\figref{length-width}.
\end{itemize}
When limiting these parameters to certain values, we do not extract longer or wider paths. We tune the optimal values of width and length by grid search of combinations on a validation set of programs and choose the combination that yields the highest accuracy, as described in~\secref{evaluation}. The tuning process of finding the optimal parameter values should be separate for each language and task.

Obviously, setting the values of these parameters to a value that is too low limits the expressiveness of the paths, does not capture enough context for each element, limits the ability to model the training and test data, and therefore produces poor accuracy. Why, then, does limiting the path length and width actually improve accuracy? There are several reasons:
\begin{itemize}
\item {\bfseries Locality} The role of a program element is affected mostly by its surroundings. For example, consider the program in~\figref{widthExample}. The width of a path that connects the variable \scode{a} in the first line to the variable \scode{b} in the last line is as large as the number of lines in the program. Usually, the names of \scode{a} and \scode{b} can be predicted by considering elements within a closer distance. Therefore, using paths between too distant elements can cause noise and pollute the relevant information.
\item {\bfseries Sparsity} Using paths that are too long can cause the representation space to be too sparse. A long path might appear too few times (or even only once) in the training set and cause the model to predict specific labels with high probability. This phenomenon is known as \emph{overfitting}, where the learned AST paths are very specific to the training data and the model fails to generalize to new, unseen data.
\item {\bfseries Performance} There is a practical limit on the amount of data that a model can be trained on. Too much data can cause the training phase to become infeasibly long. There is a tradeoff between how many programs the model can be trained on, and how many paths are extracted from each program. Therefore, it makes sense to limit the number of extracted paths from each program by limiting the paths' length and width, in order to be able to train on a larger and more \emph{varied} training set. 
\end{itemize}

\begin{figure}
\centering
\lstset{language=JavaScript, basicstyle=\footnotesize\ttfamily,emph={fence},emphstyle=\underbar,escapeinside={(*@}{@*)}}
\begin{tabular}{cc}
\begin{lstlisting}
assert.equal(a,1);
assert.equal(...);
...
assert.equal(b,1);
\end{lstlisting}
\end{tabular}
\caption{An example of a typical program where the maximal path length is relatively small, but the width can be large.}
\label{Fi:widthExample}
\end{figure}

In fact, tuning path length and width is used to control the bias-variance tradeoff. Shorter paths increase the bias error, while longer paths increase the variance error. The relationship between these parameters and results is discussed and demonstrated in~\secref{hyperparams}.

%% file: statementpath-figure.tex
\begin{figure}
\begin{subfigure}[b]{0.2\textwidth}
\centering
\lstset{language=JavaScript, basicstyle=\footnotesize\ttfamily,emph={fence},emphstyle=\underbar,showlines=true,escapeinside={(*@}{@*)}}

\begin{lstlisting}

var item = array[i];

\end{lstlisting}
\caption{}
\label{Fi:childIdExampleLine}
\end{subfigure}
\unskip \vrule
\hspace{1mm}
\begin{subfigure}[b]{0.25\textwidth}
\centering
\includegraphics[width=1.8in]{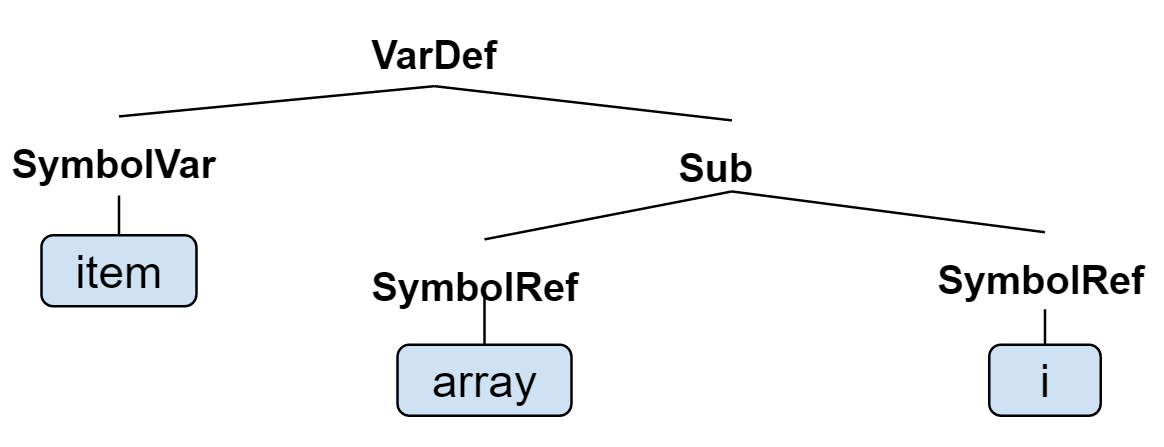}
\caption{}
\label{Fi:statementPath}
\end{subfigure}
\caption{A JavaScript statement and its partial AST.}
\label{childIdExample}
\end{figure}

%% file: evaluation.tex
\section{Evaluation}\seclabel{evaluation}

Since the goal of this work is to \emph{provide a representation of program elements}, we compared the effect of different representations on the accuracy of the learning algorithms.
To show that our approach can be applied to the representation of the input without modifying the learning algorithm, we used off-the-shelf learning algorithms but represented the input in each experiment using \emph{a different representation} (when possible).

Our evaluation aims to answer the following questions:
\begin{itemize}
\item How useful are AST paths compared to existing representations? (\secref{results})
\item How useful are AST paths across different programming languages, tasks and learning algorithms? (\secref{results})
\item Do AST paths just memorize the input, or do they capture deeper semantic regularities? (\secref{qualitative})
\item How long are the useful paths? How do the paths' length and width affect the results? (\secref{hyperparams})
\item How important is the concrete representation of paths? Which abstractions can be used to represent paths without reducing accuracy? (\secref{Abstractions})
\end{itemize}

\para{Leafwise and semi-paths}
Although the family of representations in this work includes n-wise paths and paths between any kind of AST nodes, for simplicity and feasible training time, we performed most of the experiments using leafwise-paths (paths between AST terminals) and semi-paths --- paths between an AST terminal and one of its ancestor nodes in the AST. The idea is that leafwise-paths are more diverse and therefore more expressive than semi-paths, but semi-paths provide more \emph{generalization}. Semi-paths allow us to generalize learning and capture common patterns in different programs, even if the full path does not recur.

An exception is the prediction of full types in Java, in which we predict types of expressions which are not necessarily terminals. In this case, we also used paths between terminals to the nonterminal in question.

\subsection{Prototype Implementation}\seclabel{implementation}

We implemented a cross-language tool called \tool. The tool consists of separate modules that parse and traverse the AST of a program in each different language, but the main algorithm is the same across all languages. Currently \tool contains modules for Java, JavaScript, Python and C\#, and it can be easily extended to any other language.

\para{AST construction and path extraction}
For Java we used~\citetalias{JavaParser}; for JavaScript we used~\citetalias{UglifyJS} for parsing and traversing the AST, along with additional modifications from~\citetalias{UnuglifyJS}; for Python we used the Python internal parser and AST visitor; and for C\# we used~\citetalias{Roslyn}.

\para{Learning algorithms} We experiment with two learning algorithms: Conditional Random Fields, based on the implementation of Nice2Predict~ \cite{jsnice2015}, and the word2vec based implementation of \citet{levy2014dependency}.

To support our representation in the learning engine side and produce a qualitative evaluation, we introduced minor extensions to the Nice2Predict framework:
\begin{itemize}
\item \emph{Support unary factors.} Previously, Nice2Predict supported only pairwise feature functions, and we implemented support for unary factors to express the relationship between different occurrences of the same identifier. Note that this is required because different AST nodes for the same identifier are merged into a single node in the CRF. Hence, a path between these nodes in the AST becomes a unary-factor in the CRF. This extension increases accuracy by about 1.5\%.
\item \emph{Top-k candidates suggestion.} CRFs output a single prediction for each program element. We implemented an additional API that receives a parameter $k$ and suggests the \emph{top-k} candidate names for each program element (this extension was adopted into Nice2Predict).
    This allowed us to manually investigate the quality of results (\secref{qualitative}). When all top-k predictions for a variable name captured similar notions, it increased our confidence that the model performs stable predictions.
\end{itemize}

\subsection{Experimental Setting}\seclabel{expsetting}

\para{Data sets}
For each language, we collected source code from public GitHub projects, and split it randomly to training, validation and test sets. Our data included the top ranked projects of each language and the projects that were forked the most. \tabref{datasets} shows the amount of data used for each language. Java required an order of magnitude more data than the other languages: we had to keep enlarging our Java dataset to achieve results that were close to the other languages.

Following recent work which found a large amount of code duplication in GitHub~\cite{dejavu2017}, we devoted much effort into filtering duplicates from our dataset, and especially the JavaScript dataset. To filter duplicates, we used file names, directory names (such as ``node\_modules''), and md5 of files. In Java and Python, which do not commit dependencies, duplication is less severe (as also observed by \citet{dejavu2017}). Furthermore, in our setting, we took the top-ranked and most popular projects, in which we observed duplication to be less of a problem (\citet{dejavu2017} measured duplication across \emph{all} the code in GitHub).

\para{Evaluation metric} For simplicity, in all the experiments we measured the percentage of exact match predictions, case-insensitive and ignoring differences in non-alphabetical characters. For example, this metric considers \scode{totalCount} as an exact match to \scode{total\_count}. An exception is the comparison to \citet{conv16}, who optimized their Java method name prediction model to maximize the F1 score over sub-tokens. In this case, we compared their model with ours on both exact match and F1 score. An unknown test label (``UNK'') was always counted as an incorrect prediction, or as a possibly partial prediction when using the F1 score, and our model never suggests ``UNK''. For example, if the true test label is \scode{get<UNK>}, our model could get partial precision and partial recall for predicting \scode{getFoo}.

\begin{table}[]
\centering
\footnotesize
\begin{tabular}{lrrrrrrr}
\toprule
                    & Total &\multicolumn{2}{c}{Training Set}  & \multicolumn{2}{c}{Test set} \\
Language            & repos & File   & Size (GB)    & File & Size (MB) \\ 
\midrule
Java                & $10,081$             & $1,717,016$        & $16$         & $50,000$        & 1001  \\ 
JavaScript          & $6,863$             & $159,038$          & $3.4$        & $38,103$        & 130  \\ 
Python              & $8,565$             & $458,771$          & $5.4$        & $39,941$         & 588 \\ 
C\#                 & $1,000$             & $262,774$          & $4.7$        & $50,000$        & 1208   \\ 
\bottomrule
\end{tabular}
\caption{The amounts of data used for the experimental evaluation of each language.}
\tablabel{datasets}
\end{table}

\subsection{Quantitative Evaluation}\seclabel{results}

\input{results-table}
\input{results-nn-table}

We conducted several experiments to evaluate the usefulness of AST paths in different tasks and programming languages.
We performed the following quantitative experiments:
\begin{itemize}
	\item \emph{Prediction of variable names across all four languages}. Variable names have sufficient training data in all languages to produce meaningful results. In this experiment we used both CRFs and word2vec. As baselines we used the work of \citet{jsnice2015}, CRFs with token-based n-grams as factors, and a simple rule-based baseline. For JavaScript with word2vec, we used word2vec with linear token context as a baseline and show that path representations yield dramatic improvement.
	 	
	\item \emph{Prediction of method names across JavaScript, Java and Python}. We compared our general approach for method name prediction with \citet{conv16}, who used a convolutional neural network with attention.

	
	\item \emph{Prediction of full types in Java}. For Java, we compared our results to a synthetic (straw-man) baseline that predicts all types to be \scode{java.lang.String}. This baseline shows that despite the prevalence of the \scode{String} type, the task of type prediction is still very challenging. 
\end{itemize}

In all of the following CRF experimental setups, ``no-path'' refers to a ``bag-of-words'' baseline, in which we used the same CRF learning algorithm, but used a single symbol to represent all relations. In this baseline, path information was hidden from the model during training and testing, and therefore it always assigned the same likelihood for each specific pair of identifiers, regardless of the syntactic relation between them. This baseline can be seen as a ``bag of near identifiers'' that uses the neighbors' names without their syntactic relation and therefore without considering the \emph{way} program elements are related.

\subsubsection{Predicting Variable Names}
To predict variable names, we used both CRFs and word2vec.

\para{Evaluation with CRFs} We present our evaluation results with CRFs for names in the top part of~\tabref{results}.
For JavaScript, where a tool that uses predefined features exists, we evaluated the other tool with the exact same datasets and settings, and the same AST terminals as CRF nodes, which makes the input representation (AST paths vs. their features) the only difference between the two experiments.
Using our representations yields $7.6\%$ higher accuracy than the previous work.

For Java, we compared the results with two baselines:
\begin{itemize}
\item \emph{CRFs $+$ n-grams} - this baseline uses the same CRF nodes as the path-based model, except that the relations between them are the \emph{sequential} n-grams. We chose $n=4$ as the value that maximizes accuracy on the validation set, such that the produced model consumes approximately the same amount of memory and disk as the path-based model.
\item \emph{Rule-based} - Since Java is a typed language which has a rich type system, and typical code tends to use a lot of classes and interfaces, we wonder whether the task of predicting variable names is easier in Java than in other languages and can be solved using traditional rule-based (non-learning) approaches. Our rule-based baseline predicts variable names based on the following pattern heuristics and statistics of the training corpus:
    \begin{itemize}
    \item \scode{for(int \underline{i} = ...) \{}
    \item \scode{this.<fieldName> = \underline{<fieldName>};}
    \item \scode{catch (... \underline{e}) \{}
    \item \scode{void set<fieldName>(... \underline{<fieldName>}) \{}
    \item Otherwise: use the type: \scode{HttpClient \underline{client}}.
    \end{itemize}
\end{itemize}

As shown, using CRFs with AST paths yields higher results than the baselines, in all the languages, showing that our representation yields higher results than manually defined features, n-grams, and rule-based approaches.

\para{Evaluation with word2vec} We present our evaluation results with a word2vec based implementation in~\tabref{results-nn}. For comparison, we use two alternative approaches to represent the context for prediction:
\begin{itemize}
\item The \emph{linear token-stream} approach uses the surrounding tokens to predict a variable name. Surrounding tokens (e.g., values, keywords, parentheses, dots and brackets) may implicitly hint at the syntactic relations, without AST paths. This is the type of context usually used in NLP, in the original implementation of word2vec, and in many works in programming languages.
\item The \emph{path-neighbors, no-paths} approach uses the same surrounding AST nodes for contexts as AST paths, except that the path itself is hidden, and only the identity of the surrounding AST nodes is used. The goal of using this baseline is to show that the advantage of AST paths over token-stream is not only in their wider \emph{span}, but in the representation of the path itself.
\end{itemize}

Using word2vec with AST paths produces much better results compared to these baselines. This shows the advantage of using AST paths as context over token-stream based contexts, and the significance of using a representation of the paths for prediction.

\para{Limitations of evaluation}
We noticed that our models often predict names that are very similar but not identical to the original name, such as \scode{message} instead of \scode{msg}, or synonyms such as \scode{complete} instead of \scode{done}; these are counted as incorrect predictions. Moreover, we noticed that our models sometimes predict \emph{better} names than the original names. Therefore, the accuracy results are an underapproximation of the ability of AST paths to predict meaningful names.

Another limitation lies in the inability of CRFs and word2vec to predict out-of-vocabulary (OoV) names. As was previously observed~\cite{conv16, allamanis2015}, there are two main types of OoV names in programs: names that did not appear in the training corpus but can be composed of known names (neologisms), and entirely new names. The total OoV rate among our various datasets and tasks varied between $5-15\%$, and specifically $7\%$ for predicting variable names in JavaScript, and $13\%$ for Java method names. Several techniques were suggested to deal with each type of OoV~\cite{conv16, allamanis2015}, which we did not consider here and are out of scope of this work.

\para{Discussion} We note that the accuracy for Java is lower than for JavaScript. We have a few possible explanations:
\begin{inparaenum}[(i)]
	\item The JavaScript training set contains projects that are rather domain specific, mostly client and server code for web systems (for example, the terms \scode{request} and \scode{response} are widely used across all projects). In contrast, the Java code is much more varied in terms of domains.
	\item The Java naming scheme makes extensive use of compound names (e.g., \scode{multithreadedHttpConnectionManager}), and this is amplified by the type-based name suggestions for variables provided by modern Java IDEs. In contrast, the JavaScript variable names are typically shorter and are not an amalgamation of multiple words (e.g., \scode{value}, \scode{name}, \scode{elem}, \scode{data} are frequent names).
\end{inparaenum}

The accuracy of C\# is similar to Java, but using significantly less training data. We believe that C\# naming is more structured because the commonly used C\# IDE (VisualStudio), suggests variable names based on their types.

The accuracy for Python is lower than that of JavaScript. Manual examination of the training data shows that Python programs vary widely in code quality, making the training set more noisy than that of other languages. In addition, the variety of domains and IDEs for Python makes variable names less standard. Finally, Python is easy to write, even for non-programmers, and thus there is a wide variety of non-professional Python code. The low accuracy for Python is also consistent with \citet{decisionTrees2016}.

\para{Comparison of CRFs and word2vec}
We observe that the accuracy of \tool + CRFs is higher than that of \tool + word2vec, as can be seen in~\tabref{results}.
One reason is that, unlike CRFs, word2vec was not designed exactly for this prediction task. Originally, word2vec was intended to produce meaningful word embeddings: given a set of query path-contexts, the vectors of all of them are assigned the same weight for predicting the unknown value.

Moreover, CRFs are relatively more interpretable. The weights assigned to each factor can be observed and explain a prediction posteriori.
However, word2vec was faster to train and much more memory efficient. In our evaluation, the memory required for training was over $200$GB for CRFs and only $10$GB with word2vec. Further, the training time of CRFs was up to $100$ hours, where word2vec required at most $5$ hours.


The goal here is not to provide a fair comparison between CRFs and word2vec, as their prediction tasks are slightly different; our observations in this regard are merely anecdotal. The main goal is to compare \emph{different representations for the same learning algorithm} and show that each of the learning algorithms separately can be improved by plugging in our simple representation.

\subsubsection{Predicting Method Names}
We present our evaluation results for predicting method names in~\tabref{results}. Accuracy was similar for all languages ($\sim50\%$).

Good method names balance the need to describe the internal implementation of the method and its external usage~\cite{Host:2009:DMN:1615184.1615204}. For predicting method names, we use mostly the paths from within a method to its name, but when available in the same file, we also use paths from invocations of the method to the method name. Ideally, one would use paths from different files (and for library methods, even across projects), but this requires a non-local view, which we would like to avoid for efficiency reasons.

We use the internal paths from the leaf that represents the method name to other leaves within the method AST (which capture the method implementation) and the external paths from references of the method to their surrounding leaves (which represent the usage of the method). However, we observed that using only internal paths yields only $1\%$ lower accuracy.

In Java, CRFs with AST paths are compared to the model of \citet{conv16}, which we trained on the same training corpus. Since their model is optimized to maximize the F1 score over sub-tokens, \tabref{results} presents both exact accuracy and F1 score for method name prediction in Java. The table shows that CRFs with AST paths significantly improve over the previous work in both metrics.

\input{python_example}
\input{js_example}
\input{java_example}

\subsubsection{Predicting Full Types}
\begin{sloppypar}
Our results for predicting full types in Java using CRFs are shown in the bottom part of~\tabref{results}. Our goal is to predict the full type even when it explicitly appears in the code (e.g., \scode{com.mysql.jdbc.Connection}, rather than \scode{org.apache.http.Connection}). Here we also use paths from leaves to nonterminals which represent expressions. The evaluated types were only those that could be solved by a global type inference engine. Therefore, accuracy is the percent of correct predictions out of the results that are given by type inference.
\end{sloppypar}

Although a type inference engine still produces more accurate results than our learning approach, our results using AST paths are surprisingly good, especially considering the relative simplicity of our representation. We also note that type inference is a global task, and our approach reconstructs types locally without considering the global scope of the project.

CRFs with AST paths achieved $69.1\%$ accuracy when predicting full type for Java. We contrast this result with a na\"ive baseline that uniformly predicts the type \scode{java.lang.String} for all expressions. This naive baseline yields an accuracy of $24.1\%$, which shows the task is nontrivial, even when factoring out the most commonly used Java type.

\subsection{Qualitative Evaluation}\seclabel{qualitative}

Our qualitative evaluation includes:
\begin{itemize}
	\item An anecdotal study of name prediction in different languages. For JavaScript we also compared our predictions to those of \citet{jsnice2015} in interesting cases.
	\item An anecdotal study of top-k predictions for some examples, showing semantic similarities between predicted names as captured by the trained model.
\end{itemize}

\subsubsection{Prediction Examples}
\figref{pythonExample_p} shows an example of a Python program predicted using AST paths. It can be seen that all the names predicted using AST paths were renamed with meaningful names such as \scode{process}, \scode{cmd} and \scode{retcode}.

\begin{sloppypar}
\figref{jsExample} shows the default JavaScript example from {\small nice2predict.org}, predicted using AST paths and an online version of UnuglifyJS at {\small nice2predict.org}. We note that their online model was not trained on the same dataset as our model. The model which was trained using UnuglifyJS and our dataset yielded worse results. It can be seen that our model produced more meaningful names such as \scode{url} (instead of \scode{source}) and \scode{callback} (instead of \scode{n}).
\end{sloppypar}

\figref{javaExample} shows examples of Java programs. To demonstrate the expressiveness of AST paths, we deliberately selected challenging examples in which the prediction cannot be aided by the informative class and interface names that Java code usually contains (as in: \scode{HttpClient \underline{client}}). Instead, our model had to leverage the syntactic structure to predict the meaningful names: \scode{done}, \scode{values}, \scode{value} and \scode{count}.

\subsubsection{Semantic Similarity between Names}

\input{semsim-table}

It is interesting to observe the other \emph{candidates} that our trained model produces for program elements. As~\tabref{doneCandidates} shows, the next candidates after \scode{done} (in~\figref{code-example}) are: \scode{ended}, \scode{complete}, \scode{found}, \scode{finished}, \scode{stop} and \scode{end}, which are all semantically similar (in programs, not necessarily in natural language).
In many cases, AST paths capture the \emph{semantic similarity} between names, for example \scode{req}$\sim$\scode{request} and \scode{list}$\sim$\scode{array}, as shown in~\tabref{topk-similarities}. This supports our hypothesis that AST paths capture the semantic role of a program element.

\subsection{Impact of Parameter Values}\seclabel{hyperparams}

In~\secref{application} we introduced and discussed the importance of the \maxlen and \maxwidth parameters. For each language we experimented with different combinations of values for \maxlen and \maxwidth on its validation set. We chose the values that produced the highest accuracy while still being computationally feasible when evaluating the model with the test set.

\para{Accuracy with different path length and width} We experimented with tuning the path parameters and observed their effect on the accuracy. The best parameter values for each prediction are shown in~\tabref{results}.

\input{accuracy-vs-len-figure}

For the task of name prediction, for all languages, the best path length is 6-7, and the best width is 3-4. The variations in path length stem from minor differences in the structure of the AST. For example, despite the similarity in source level between Java and C\#, the C\# AST is slightly more elaborate than the one we used for Java.

A drill-down of the accuracy given different parameter values for variable name prediction in JavaScript is shown in~\figref{accuracyvslength}. We observe that the \maxlen parameter has a significant positive effect, while the contribution of a larger \maxwidth is positive but minor. This observation affirms our initial hypothesis that our long-distance paths are fundamental and crucial to the accuracy of the prediction. It also confirms our belief that an automatic representation of code (rather than manually defined) is essential, since the long-distance paths are very unlikely to have been designed manually.

For the task of method name prediction, since there are significantly fewer paths, we could afford to set a high parameter value without too much tuning and still keep the training time and resources feasible. We therefore set the length in this case to $12$ for JavaScript, $10$ for Python, and just $6$ for Java.

For the task of predicting full types in Java, we used length $4$ and width $1$, which yielded an accuracy of $69.1\%$. The intuition for the short path length is that in many cases the type of an expression can be inferred locally from other neighboring types, often from an explicit type declaration.

Higher values for \maxlen and \maxwidth resulted in higher training times, but combined with the \emph{downsampling} approach, it is possible to maintain a shorter training time while increasing the parameter values, and control the tradeoff between accuracy and training time.

\input{random-pigeon-figure}

\para{Downsampling} We wanted to measure the effect of training data size on accuracy and training time. Inspired by \citet{node2vec2016}, who used random walks to learn representations for neighborhoods of nodes in a network, we experimented with randomly omitting a fraction of the extracted path-contexts. After extracting path-contexts from all the programs in the training set, we randomly omitted each occurrence of a path-context in probability of $1-p$ (i.e., $p$ is the probability to keep it) and trained a model only on the remaining paths. As can be seen in~\figref{randompigeon}, randomly dropping contexts can significantly reduce training time, with a minimal effect on accuracy. 
For example, for $p=0.8$ we observed exactly the same accuracy as for the complete set of paths ($p=1$), while training time was reduced by about $25\%$. Moreover, decreasing $p$ down to $0.2$ still yielded higher accuracy than UnuglifyJS but reduced training time by about $80\%$ (compared to $p=1.0$).

\subsection{Abstractions of AST Paths}\seclabel{Abstractions}
In order to evaluate the full expressiveness of AST paths, the previously reported experiments were performed using no abstraction, i.e. $\alpha_{id}$. However, it is also possible to use a higher level of abstraction. Instead of representing the whole path node-by-node with separating up and down arrows, it is possible to keep only parts of this representation. This abstraction results in less expressive paths and might represent two different paths as equal, but it enables decreasing the number of distinct paths, thus reducing the number of model parameters. Training will be faster as a result.

Different levels of path abstractions also allow us to evaluate the importance of different components of AST paths, or \emph{which components of AST paths contribute to their usefulness the most}. We experimented with several levels of abstraction:
\begin{itemize}
\item ``No-arrows'' - using the full path encoding, except the up and down symbols $\left\{\uparrow ,\downarrow\right\}$.
\item ``Forget-order'' - using paths without arrows and without order between the nodes: instead of treating a path as a \emph{sequence} of nodes, treat it as a \emph{bag} of nodes.
\item ``First-top-last'' - keeping only the first, top and last nodes of the path. The ``top'' node refers to the node that is hierarchically the highest, from which the direction of the path changes from upwards to downwards.
\item ``First-last'' - keeping only the first and last nodes.
\item ``Top'' - keeping only the top node.
\item ``No-paths'' - using no paths at all, and treating all relations between program elements as the same. The name of an element is predicted by using the bag of surrounding identifiers, without considering the syntactic relation to each of them.
\end{itemize}

All of the following experiments were performed using CRFs for variable names prediction, on the Java corpus and on the same hardware. In every experiment, the training corpus and the rest of the settings were identical. The number of training iterations was fixed.

\figref{abstractions_figure} shows the accuracy of each abstraction compared to the consumed training time. As shown, as more information is kept, accuracy is increased, with the cost of a longer training time. An interesting ``sweet-spot'' is ``first-top-last'', which reduces training time by half compared to the full representation, with accuracy that is as $95\%$ as good.

We also observe that the arrows and the order of the nodes in the path contribute about $1\%$ accuracy.

\input{abstractions_figure} 

%% file: results-table.tex
\begin{table*}[]
\centering
\footnotesize
\begin{tabular}{llllr}
\toprule
Task              & \multicolumn{2}{c}{Previous works} & AST Paths (this work)   & Params (length/width)   \\ 
\midrule
\textbf{Variable name prediction} &                             &        &               \\
\tab JavaScript               & $24.9\%$ (no-paths)   & $60.0\%$ (UnuglifyJS) & \textbf{\jsacc} &  $7$/$3$          \\
\tab Java                     & $23.7\%$ (rule-based) & $50.1\%$ (CRFs$+4$-grams)& \textbf{\javaacc}               &  $6$/$3$          \\
\tab Python                   & $35.2\%$ (no-paths)   &               & \textbf{\pythonacc}             &  $7$/$4$          \\
\tab C\#                      &                       &               & \textbf{56.1\%}                 &  $7$/$4$          \\  \\
\textbf{Method name prediction} &                     &               &                        &               \\
\tab JavaScript               & $44.1\%$ (no-paths)   &               & \textbf{53.1\%}                 &  $12$/$4$         \\
\tab Java                     & \multicolumn{2}{l}{$16.5\%$, F1: $33.9$ (\citet{conv16})}             & \textbf{47.3\%}, F1: \textbf{49.9}                 &  $6$/$2$          \\
\tab Python                   & $41.6\%$ (no-paths)   &               & \textbf{51.1\%}                 &  $10$/$6$         \\  \\
\textbf{Full type prediction} &                       &               &                        &               \\
\tab Java                     & $24.1\%$ (na\"ive baseline)&          & \textbf{69.1}\% & $4$/$1$  \\
\bottomrule
\end{tabular}
\caption{Accuracy comparison for variable name prediction, method name prediction, and full type prediction using CRFs. }
\tablabel{results}
\end{table*} 

%% file: results-nn-table.tex
\begin{table}[]
\centering
\footnotesize
\begin{tabular}{ll}
\toprule
\multicolumn{1}{l}{Model} & \multicolumn{1}{l}{Names Accuracy} \\ 
\midrule
linear token-stream + word2vec    & $20.6\%$                              \\
path-neighbors, no-paths + word2vec        & $23.2\%$                              \\
\textbf{AST Paths (this work) + word2vec }                       & \textbf{40.4\%}                           \\ 
\bottomrule
\end{tabular}
\caption{Accuracy comparison for the variable name prediction task that was evaluated using word2vec in JavaScript.}
\tablabel{results-nn}
\end{table} 

%% file: python_example.tex
\begin{figure*}[]
\centering
\begin{tabular}{|l|l|}
\hline
Stripped names & AST paths + CRFs \\ \hline
\begin{subfigure}[b]{0.33\textwidth}
\tikz[remember picture]\node[inner sep=0pt,outer sep=0pt] (a){
\lstset{language=Java, basicstyle=\scriptsize\ttfamily,emph={c,p,o,e,r},emphstyle=\bfseries,escapeinside={(*}{*)},morekeywords={def,for,in}}
\begin{lstlisting}
def sh3(c):
    p = Popen(c, stdout=PIPE,
        stderr=PIPE, shell=True)
    o, e = p.communicate()
    r = p.returncode
    if r:
        raise CalledProcessError(r, c)
    else:
        return o.rstrip(), e.rstrip()
\end{lstlisting}};
\end{subfigure}
         &
\begin{subfigure}[b]{0.39\textwidth}
\tikz[remember picture]\node[inner sep=0pt,outer sep=0pt] (b){
\lstset{language=Java, basicstyle=\scriptsize\ttfamily,emph={cmd,process,out,err,retcode},emphstyle=\bfseries,escapeinside={(*}{*)},morekeywords={def,for,in}}

\begin{lstlisting}
def sh3(cmd):
    process = Popen(cmd, stdout=PIPE,
        stderr=PIPE, shell=True)
    out, err = process.communicate()
    retcode = process.returncode
    if retcode:
        raise CalledProcessError(retcode, cmd)
    else:
        return out.rstrip(), err.rstrip()
\end{lstlisting}};
\end{subfigure}
\\ \hline
\end{tabular}
\caption{Example of a Python program with stripped names and with predictions produced using our AST paths.}
\label{Fi:pythonExample_p}
\end{figure*} 

%% file: js_example.tex
\begin{figure*}[]
\centering
\begin{tabular}{|l|l|l|}
\hline
Stripped Names & AST Paths + CRFs & \url{nice2predict.org}  \\ \hline
\begin{subfigure}[b]{0.3\textwidth}
\tikz[remember picture]\node[inner sep=0pt,outer sep=0pt] (e){
\lstset{language=JavaScript, basicstyle=\scriptsize\ttfamily,emphstyle=\underbar,escapeinside={(*}{*)}}
\begin{lstlisting}
function f((*\bfseries a*), (*\bfseries b*), (*\bfseries c*)) {
  (*\bfseries b*).open('GET', (*\bfseries a*), false);
  (*\bfseries b*).send((*\bfseries c*));
}
\end{lstlisting}};
\end{subfigure}
&
\begin{subfigure}[b]{0.3\textwidth}
\tikz[remember picture]\node[inner sep=0pt,outer sep=0pt] (f){
\lstset{language=JavaScript, basicstyle=\scriptsize\ttfamily,emphstyle=\underbar,escapeinside={(*}{*)}}

\begin{lstlisting}
function f((*\bfseries url*), (*\bfseries request*), (*\bfseries callback*)) {
  (*\bfseries request*).open('GET', (*\bfseries url*), false);
  (*\bfseries request*).send((*\bfseries callback*));
}
\end{lstlisting}};
\end{subfigure}
&
\begin{subfigure}[b]{0.3\textwidth}
\tikz[remember picture]\node[inner sep=0pt,outer sep=0pt] (i){
\lstset{language=JavaScript, basicstyle=\scriptsize\ttfamily,emphstyle=\underbar,escapeinside={(*}{*)}}

\begin{lstlisting}
function f((*\bfseries source*), (*\bfseries req*), (*\bfseries n*)) {
  (*\bfseries req*).open("GET", (*\bfseries source*), false);
  (*\bfseries req*).send((*\bfseries n*));
}
\end{lstlisting}};
\end{subfigure}            \\ \hline
\end{tabular}
\caption{Example of a JavaScript program with stripped names, with predictions produced using our AST paths and an online version of UnuglifyJS at {\small nice2predict.org}. This is the default example shown at {\small nice2predict.org}.}
\figlabel{jsExample}
\end{figure*} 

%% file: java_example.tex
\begin{figure*}[]
\centering
\begin{tabular}{|l|l|}
\hline
Stripped names & AST paths + CRFs \\ \hline

\begin{subfigure}[b]{0.33\textwidth}
\tikz[remember picture]\node[inner sep=0pt,outer sep=0pt] (a){
\lstset{language=Java, basicstyle=\scriptsize\ttfamily,emphstyle=\underbar,escapeinside={(*}{*)}}

\begin{lstlisting}
boolean (*\bfseries d*) = false;
while (!(*\bfseries d*)) {
    if (someCondition()) {
        (*\bfseries d*) = true;
    }
}
\end{lstlisting}};
\end{subfigure}
         &
         \begin{subfigure}[b]{0.39\textwidth}
\tikz[remember picture]\node[inner sep=0pt,outer sep=0pt] (b){
\lstset{language=Java, basicstyle=\scriptsize\ttfamily,emphstyle=\underbar,escapeinside={(*}{*)}}

\begin{lstlisting}
boolean (*\bfseries done*) = false;
while (!(*\bfseries done*)) {
    if (someCondition()) {
        (*\bfseries done*) = true;
    }
}
\end{lstlisting}};
\end{subfigure}
\\ \hline
\begin{subfigure}[b]{0.33\textwidth}
\tikz[remember picture]\node[inner sep=0pt,outer sep=0pt] (c){
\lstset{language=Java, basicstyle=\scriptsize\ttfamily,emphstyle=\underbar,escapeinside={(*}{*)}}

\begin{lstlisting}
int count(List<Integer> (*\bfseries x*), int (*\bfseries t*)) {
    int (*\bfseries c*) = 0;
    for (int (*\bfseries r*): (*\bfseries x*)) {
        if ((*\bfseries r*) == (*\bfseries t*)) {
            (*\bfseries c*)++;
        }
    }
    return (*\bfseries c*);
}
\end{lstlisting}};
\end{subfigure}
         &
         \begin{subfigure}[b]{0.39\textwidth}
\tikz[remember picture]\node[inner sep=0pt,outer sep=0pt] (d){
\lstset{language=Java, basicstyle=\scriptsize\ttfamily,emphstyle=\underbar,escapeinside={(*}{*)}}
\begin{lstlisting}
int count(List<Integer> (*\bfseries values*), int (*\bfseries value*)) {
    int (*\bfseries count*) = 0;
    for (int (*\bfseries v*): (*\bfseries values*)) {
        if ((*\bfseries v*) == (*\bfseries value*)) {
            (*\bfseries count*)++;
        }
    }
    return (*\bfseries count*);
}
\end{lstlisting}};
\end{subfigure}
\\ \hline
\end{tabular}
\caption{Examples of Java programs with stripped names and with predictions produced using our AST paths. We deliberately selected challenging examples in which the prediction cannot be aided by specific classes and interfaces.}
\figlabel{javaExample}
\end{figure*}

%% file: semsim-table.tex
\begin{table}[]
\centering
\footnotesize
\begin{tabular}{c}
\begin{subtable}	[t]{.7\linewidth}
\centering
\begin{tabular}{ll}
\toprule
   & Candidate  \\ 
\midrule
1. & done           \\ 
2. & ended        \\ 
3. & complete            \\ 
4. & found      \\ 
5. & finished            \\ 
6. & stop          \\ 
7. & end             \\ 
8. & success        \\ 
\bottomrule
\end{tabular}
\caption{Candidates for prediction of the variable \scode{d} from the example program of \figref{code-example}.}
\tablabel{doneCandidates}
\end{subtable}
\\
\begin{subtable}[t]{0.7\linewidth}
\begin{tabular}{l}
\toprule
Semantic Similarities                       \\ 
\midrule
req $\sim$ request $\sim$ client                                            \\
items $\sim$ values $\sim$ objects $\sim$ keys $\sim$ elements\\
array $\sim$ arr $\sim$ ary $\sim$ list\\
item $\sim$ value $\sim$ key $\sim$ target \\
element $\sim$ elem $\sim$ el \\
count $\sim$ counter $\sim$ total \\
res $\sim$ result $\sim$ ret \\
i $\sim$ j $\sim$ index \\ 
\bottomrule
\end{tabular}
\caption{Examples of semantic similarities between names found among the top-10 candidates.}
\tablabel{topk-similarities}
\end{subtable}
\end{tabular}
\caption{Semantic similarities between names.}
\end{table} 

%% file: accuracy-vs-len-figure.tex
\begin{figure}
\begin{tikzpicture}[scale=1.0]
	\begin{axis}[
		xlabel={Max path length value},
		ylabel={Accuracy (\%)},
        legend style={at={(0.97,0.2)},anchor=east,font=\tiny},
        xmin=2, xmax=8,
        ymin=50.0, ymax=68,
        xtick={3,4,5,6,7},
        ytick={50.0,52.0,...,68.0}
    ]
	
    \addplot[color=blue, mark=*] coordinates {
		(3,55.9)
		(4,63.3)
		(5,66.3)
		(6,67.3)
		(7,67.6)
	};
    \addlegendentry{\footnotesize AST Paths with \maxwidth=3}

    \addplot[color=blue, mark=square*] coordinates {
		(3,55.2)
		(4,62.8)
		(5,65.9)
		(6,67.0)
		(7,67.2)
	};
    \addlegendentry{\footnotesize AST Paths with \maxwidth=2}
    \addplot[color=blue, mark=triangle*] coordinates {
		(3,52.2)
		(4,60.8)
		(5,64.4)
		(6,65.5)
		(7,66.2)
	};
    \addlegendentry{\footnotesize AST Paths with \maxwidth=1}
    \addplot[color=red,domain=2:8] {
		60.0
	};
    \addlegendentry{\footnotesize UnuglifyJS (\citet{jsnice2015})}
	\end{axis}

\end{tikzpicture}
    \caption{Accuracy results of AST paths with CRFs, for the task of variable naming in JavaScript, for different combination values of \maxlen and \maxwidth (UnuglifyJS is presented here for comparison).}
    \figlabel{accuracyvslength}
\end{figure}
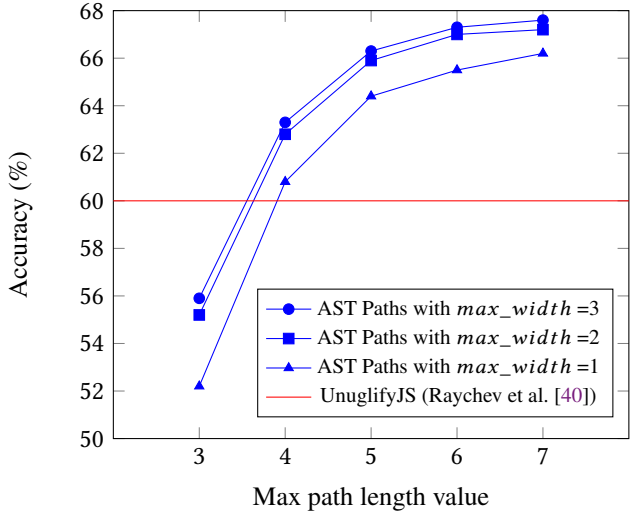 

%% file: random-pigeon-figure.tex
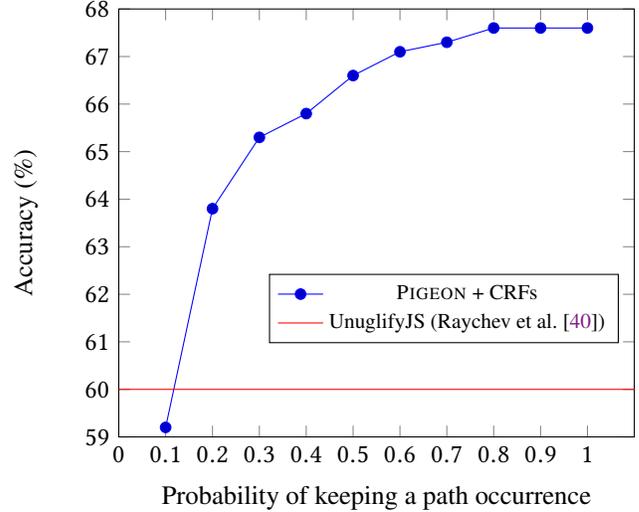
\begin{figure}
\begin{tikzpicture}[scale=1.0]
	\begin{axis}[
		xlabel={Probability of keeping a path occurrence},
		ylabel={Accuracy (\%)},
        legend style={at={(0.97,0.3)},anchor=east,font=\tiny},
        xmin=0, xmax=1.1,
        ymin=59.0, ymax=68,
        xtick={0,0.1,0.2,0.3,0.4,0.5,0.6,0.7,0.8,0.9,1.0},
        ytick={59.0,60.0, 61.0,...,68.0}
    ]
	\addplot coordinates {
		(0.1,59.2)
		(0.2,63.8)
		(0.3,65.3)
		(0.4,65.8)
		(0.5,66.6)
		(0.6,67.1)
		(0.7,67.3)
		(0.8,67.6)
		(0.9,67.6)
		(1.0,67.6)
	};
    \addlegendentry{\footnotesize \tool+ CRFs}
    \addplot[color=red,domain=0:1.1] {
		60.0
	};
    \addlegendentry{\footnotesize UnuglifyJS (\citet{jsnice2015})}
	\end{axis}

\end{tikzpicture}
    \caption{Downsampling: accuracy results of AST paths with CRFs, for the task of variable naming in JavaScript, for different values of $p$ - the probability of keeping an AST path occurrence. UnuglifyJS was evaluated with all the training data and is presented here for comparison.}
    \figlabel{randompigeon}
\end{figure}

%% file: abstractions_figure.tex
\begin{figure}
\begin{tikzpicture}[scale=1.0]
	\begin{axis}[
		xlabel={Training time (hours)},
		ylabel={Accuracy (\%)},
        legend style={at={(0.97,0.2)},anchor=east,font=\tiny},
        xmin=5, xmax=20,
        ymin=35.0, ymax=60,
        xtick={5,...,20},
        ytick={35.0,40.0,...,60.0}
    ]
	\addplot[
    blue,
    thick,
    mark=star, only marks,
    mark options={fill=white},
    visualization depends on=\thisrow{alignment} \as \alignment,
    nodes near coords, 
    point meta=explicit symbolic, 
    every node near coord/.style={anchor=\alignment} 
    ] table [
     meta index=2 
     ] {
x       y       label       alignment
5.5    38.9    no-path      180
7    50.8    first-last          160
7.4    51.6    top          -160
8    55.3    first-top-last          180
15   57.6   forget-order    70
15   57.7   no-arrows    -40
18   58.2   full        180
};

	\end{axis}

\end{tikzpicture}
    \caption{The accuracy of each abstraction method compared to the consumed training time, for the task of variable naming in Java}
    \figlabel{abstractions_figure}
\end{figure}
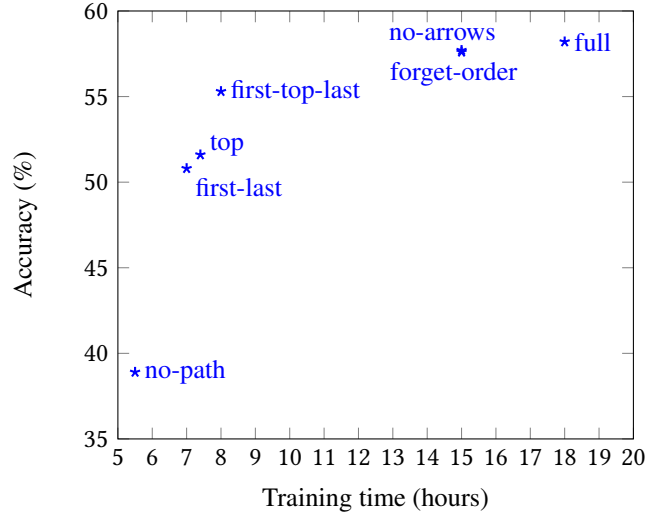 

%% file: relatedWork.tex
\section{Related Work}\seclabel{relatedWork}

\para{Naming in Software Engineering} Several studies about naming in code have been conducted \cite{takang96, butler2009, Host:2009:DMN:1615184.1615204,allamanis2015}.
Some of them applied neural network approaches for various applications.
An approach for inferring method and class names was suggested by \citet{allamanis2015}, by learning the similarity between names; however, they used features that were manually designed for a given task.
A recent work presents a convolutional attention model \cite{conv16} and evaluates it by predicting method names. In \secref{evaluation}, we show that using our general approach yields better results.

Several works have studied the use of NLP techniques in programming languages, for applications such as estimating code similarity \cite{Zilberstein:2016:LCN:2986012.2986013}, naming convention recommendations \cite{Allamanis:2014:LNC:2635868.2635883}, program synthesis \cite{Desai:2016:PSU:2884781.2884786}, translation between programming languages \cite{Karaivanov:2014:PST:2661136.2661148} and code completion \cite{raychev14, Hindle:2012:NS:2337223.2337322, Maddison:2014:SGM:3044805.3044965}. A bimodal modeling of code and natural language was suggested by \citet{bimodal15}, for tasks of code retrieval given a natural language query, and generating a natural language description given a code snippet.

A recent work presented a model that uses LSTM for code summarization, which is interpreted as generating a natural language description for given program~\cite{codenn16}. This work presents impressive results, but requires a very large amount of human-annotated data.

\ignore{
A previous work reduced the problem of code completion to NLP by using the n-gram model \cite{raychev14}. An n-gram model has also been used by \citet{Allamanis:2014:LNC:2635868.2635883} to develop a naming convention recommendation tool.
The tool trains over a specific codebase, and learns this codebase's coding conventions by learning an n-gram model over tokens, rather than words.
However, while code conventions are project-specific, our representation can be used in both project-specific and cross-project models.
}

\para{Predicting program properties using probabilistic graphical models} CRFs have been used for structured prediction of program properties in JavaScript and Android Java code \cite{jsnice2015, android2016}. The method has been shown to produce good results for prediction of names and types, by modeling programs with CRFs. \citet{jsnice2015} defined relationships between program elements using an explicit grammar specific to JavaScript. The possible relationships span only a single statement, and do not include relationships that involve conditional statements or loops. \citet{android2016} also defined several types of features and the conditions in which each of them is used. These works motivate a representation that is extractable automatically and can be applied to different languages. Instead of defining the relationships between program elements in advance, we learn them from the training data, in an automatic process that is similar for different languages. 


\para{Parse Tree Paths} An approach which resembles ours is Parse Tree Paths (PTPs) in Natural Language Processing. PTPs were mainly used in Semantic Role Labeling (SRL) --- the NLP task of identifying semantic roles such as Message, Speaker or Topic in a natural language sentence. PTPs were first suggested by \citet{gildeajurafsky2002} for automatic labeling of semantic roles, among other linguistic features. The paths were extracted from a target word to each of the constituents in question, and the method remains very popular in SRL and general NLP systems.
The rich and unambiguous structure of programming languages renders AST paths even more important as a representation of program elements than PTPs in natural language.

\para{Code completion using PCFGs} Probabilistic Context Free Grammar (PCFG) for programming languages has been used for several tasks, such as finding code idioms~\cite{allamanis2014}.
PCFGs were generalized by \citet{phog16} by learning a relative context node on which each production rule depends, allowing conditioning of production rules beyond the parent nonterminal, thus capturing richer context for each production rule.
Even though these works use paths in an AST, they differ from our work in that the path is only used to find a context node. In our work, the path itself is part of the representation, and therefore the prediction depends not only on the context node but also on the way it is related to the element in question.

An approach for learning programs from datasets with incorrect (noisy) examples was presented by \citet{raychev2016noisy}. This approach is based on sampling the dataset and synthesizing new programs, in a feedback directed loop. 

\citet{tbcnn2016} used a tree-based representation to learn snippets of code using a tree-convolutional neural network, for tasks of code category classification. Our representation differs from their mainly in that we decompose the AST into paths, which better capture data-flow properties, whereas their representation decomposes into sub-trees.

%% file: conclusion.tex
\section{Conclusion}
We presented a simple and general approach for learning from programs. The main idea is to represent a program using paths in its abstract syntax tree (AST). This allows a learning model to leverage the structured nature of source code rather than treating it as a flat sequence of tokens.

We show that this representation can be useful in a variety of programming languages and prediction tasks, and can improve the results of different learning algorithms without modifying the learning algorithm itself.

We believe that since the representation of programs using AST paths is fundamental to programming languages, it can be used in a variety of other machine learning tasks, including different applications and different learning models.

%% file: acknowledgement.tex
\begin{acks}

We would like to thank Eytan Singher for implementing the C\# module of \tool.
We also thank Miltiadis Allamanis, Veselin Raychev and Pavol Bielik for their guidance in the use of their tools in the evaluation section.

The research leading to these results has received funding from the European Union's Seventh Framework Programme (FP7) under grant agreement no. 615688-ERC- COG-PRIME. Cloud computing resources were provided by a Microsoft Azure for Research award and an AWS Cloud Credits for Research award.

\hfill \break
\hfill \break

\end{acks} 

%% file: arxiv_draft.bbl

%% file: arxiv_draft.bbl

\begin{thebibliography}{46}


\ifx \showCODEN    \undefined \def \showCODEN     #1{\unskip}     \fi
\ifx \showDOI      \undefined \def \showDOI       #1{#1}\fi
\ifx \showISBNx    \undefined \def \showISBNx     #1{\unskip}     \fi
\ifx \showISBNxiii \undefined \def \showISBNxiii  #1{\unskip}     \fi
\ifx \showISSN     \undefined \def \showISSN      #1{\unskip}     \fi
\ifx \showLCCN     \undefined \def \showLCCN      #1{\unskip}     \fi
\ifx \shownote     \undefined \def \shownote      #1{#1}          \fi
\ifx \showarticletitle \undefined \def \showarticletitle #1{#1}   \fi
\ifx \showURL      \undefined \def \showURL       {\relax}        \fi
\providecommand\bibfield[2]{#2}
\providecommand\bibinfo[2]{#2}
\providecommand\natexlab[1]{#1}
\providecommand\showeprint[2][]{arXiv:#2}

\bibitem[\protect\citeauthoryear{??}{Jav}{[n. d.]}]%
        {JavaParser}
 \bibinfo{year}{[n. d.]}\natexlab{}.
\newblock \bibinfo{title}{{JavaParser}}.
\newblock \bibinfo{howpublished}{\url{http://javaparser.org}}.
\newblock


\bibitem[\protect\citeauthoryear{??}{Ros}{[n. d.]}]%
        {Roslyn}
 \bibinfo{year}{[n. d.]}\natexlab{}.
\newblock \bibinfo{title}{{Roslyn}}.
\newblock \bibinfo{howpublished}{\url{https://github.com/dotnet/roslyn}}.
\newblock


\bibitem[\protect\citeauthoryear{??}{Ugl}{[n. d.]}]%
        {UglifyJS}
 \bibinfo{year}{[n. d.]}\natexlab{}.
\newblock \bibinfo{title}{{UglifyJS}}.
\newblock \bibinfo{howpublished}{\url{https://github.com/mishoo/UglifyJS}}.
\newblock


\bibitem[\protect\citeauthoryear{??}{Unu}{[n. d.]}]%
        {UnuglifyJS}
 \bibinfo{year}{[n. d.]}\natexlab{}.
\newblock \bibinfo{title}{{UnuglifyJS}}.
\newblock \bibinfo{howpublished}{\url{https://github.com/eth-srl/UnuglifyJS}}.
\newblock


\bibitem[\protect\citeauthoryear{Allamanis, Barr, Bird, and Sutton}{Allamanis
  et~al\mbox{.}}{2014}]%
        {Allamanis:2014:LNC:2635868.2635883}
\bibfield{author}{\bibinfo{person}{Miltiadis Allamanis},
  \bibinfo{person}{Earl~T. Barr}, \bibinfo{person}{Christian Bird}, {and}
  \bibinfo{person}{Charles Sutton}.} \bibinfo{year}{2014}\natexlab{}.
\newblock \showarticletitle{Learning Natural Coding Conventions}. In
  \bibinfo{booktitle}{{\em Proceedings of the 22Nd ACM SIGSOFT International
  Symposium on Foundations of Software Engineering}} {\em (\bibinfo{series}{FSE
  2014})}. \bibinfo{publisher}{ACM}, \bibinfo{address}{New York, NY, USA},
  \bibinfo{pages}{281--293}.
\newblock
\showISBNx{978-1-4503-3056-5}
\showDOI{%
\url{https://doi.org/10.1145/2635868.2635883}}


\bibitem[\protect\citeauthoryear{Allamanis, Barr, Bird, and Sutton}{Allamanis
  et~al\mbox{.}}{2015a}]%
        {allamanis2015}
\bibfield{author}{\bibinfo{person}{Miltiadis Allamanis},
  \bibinfo{person}{Earl~T. Barr}, \bibinfo{person}{Christian Bird}, {and}
  \bibinfo{person}{Charles Sutton}.} \bibinfo{year}{2015}\natexlab{a}.
\newblock \showarticletitle{Suggesting Accurate Method and Class Names}. In
  \bibinfo{booktitle}{{\em Proceedings of the 2015 10th Joint Meeting on
  Foundations of Software Engineering}} {\em (\bibinfo{series}{ESEC/FSE
  2015})}. \bibinfo{publisher}{ACM}, \bibinfo{address}{New York, NY, USA},
  \bibinfo{pages}{38--49}.
\newblock
\showISBNx{978-1-4503-3675-8}
\showDOI{%
\url{https://doi.org/10.1145/2786805.2786849}}


\bibitem[\protect\citeauthoryear{Allamanis, Peng, and Sutton}{Allamanis
  et~al\mbox{.}}{2016}]%
        {conv16}
\bibfield{author}{\bibinfo{person}{Miltiadis Allamanis}, \bibinfo{person}{Hao
  Peng}, {and} \bibinfo{person}{Charles~A. Sutton}.}
  \bibinfo{year}{2016}\natexlab{}.
\newblock \showarticletitle{A Convolutional Attention Network for Extreme
  Summarization of Source Code}. In \bibinfo{booktitle}{{\em Proceedings of the
  33nd International Conference on Machine Learning, {ICML} 2016, New York
  City, NY, USA, June 19-24, 2016}}. \bibinfo{pages}{2091--2100}.
\newblock
\showURL{%
\url{http://jmlr.org/proceedings/papers/v48/allamanis16.html}}


\bibitem[\protect\citeauthoryear{Allamanis and Sutton}{Allamanis and
  Sutton}{2014}]%
        {allamanis2014}
\bibfield{author}{\bibinfo{person}{Miltiadis Allamanis} {and}
  \bibinfo{person}{Charles Sutton}.} \bibinfo{year}{2014}\natexlab{}.
\newblock \showarticletitle{Mining Idioms from Source Code}. In
  \bibinfo{booktitle}{{\em Proceedings of the 22Nd ACM SIGSOFT International
  Symposium on Foundations of Software Engineering}} {\em (\bibinfo{series}{FSE
  2014})}. \bibinfo{publisher}{ACM}, \bibinfo{address}{New York, NY, USA},
  \bibinfo{pages}{472--483}.
\newblock
\showISBNx{978-1-4503-3056-5}
\showDOI{%
\url{https://doi.org/10.1145/2635868.2635901}}


\bibitem[\protect\citeauthoryear{Allamanis, Tarlow, Gordon, and Wei}{Allamanis
  et~al\mbox{.}}{2015b}]%
        {bimodal15}
\bibfield{author}{\bibinfo{person}{Miltiadis Allamanis},
  \bibinfo{person}{Daniel Tarlow}, \bibinfo{person}{Andrew~D. Gordon}, {and}
  \bibinfo{person}{Yi Wei}.} \bibinfo{year}{2015}\natexlab{b}.
\newblock \showarticletitle{{Bimodal Modelling of Source Code and Natural
  Language}}. In \bibinfo{booktitle}{{\em {Proceedings of the 32nd
  International Conference on Machine Learning}}} {\em (\bibinfo{series}{{JMLR
  Proceedings}})}, Vol.~\bibinfo{volume}{37}. \bibinfo{publisher}{{JMLR.org}},
  \bibinfo{pages}{2123--2132}.
\newblock


\bibitem[\protect\citeauthoryear{Bengio, Ducharme, Vincent, and Janvin}{Bengio
  et~al\mbox{.}}{2003}]%
        {bengio2003}
\bibfield{author}{\bibinfo{person}{Yoshua Bengio}, \bibinfo{person}{R{\'e}jean
  Ducharme}, \bibinfo{person}{Pascal Vincent}, {and} \bibinfo{person}{Christian
  Janvin}.} \bibinfo{year}{2003}\natexlab{}.
\newblock \showarticletitle{A Neural Probabilistic Language Model}.
\newblock \bibinfo{journal}{{\em J. Mach. Learn. Res.\/}}  \bibinfo{volume}{3}
  (\bibinfo{date}{March} \bibinfo{year}{2003}), \bibinfo{pages}{1137--1155}.
\newblock
\showISSN{1532-4435}
\showURL{%
\url{http://dl.acm.org/citation.cfm?id=944919.944966}}


\bibitem[\protect\citeauthoryear{Bichsel, Raychev, Tsankov, and Vechev}{Bichsel
  et~al\mbox{.}}{2016}]%
        {android2016}
\bibfield{author}{\bibinfo{person}{Benjamin Bichsel}, \bibinfo{person}{Veselin
  Raychev}, \bibinfo{person}{Petar Tsankov}, {and} \bibinfo{person}{Martin
  Vechev}.} \bibinfo{year}{2016}\natexlab{}.
\newblock \showarticletitle{Statistical Deobfuscation of Android Applications}.
  In \bibinfo{booktitle}{{\em Proceedings of the 2016 ACM SIGSAC Conference on
  Computer and Communications Security}} {\em (\bibinfo{series}{CCS '16})}.
  \bibinfo{publisher}{ACM}, \bibinfo{address}{New York, NY, USA},
  \bibinfo{pages}{343--355}.
\newblock
\showISBNx{978-1-4503-4139-4}
\showDOI{%
\url{https://doi.org/10.1145/2976749.2978422}}


\bibitem[\protect\citeauthoryear{Bielik, Raychev, and Vechev}{Bielik
  et~al\mbox{.}}{2016}]%
        {phog16}
\bibfield{author}{\bibinfo{person}{Pavol Bielik}, \bibinfo{person}{Veselin
  Raychev}, {and} \bibinfo{person}{Martin~T. Vechev}.}
  \bibinfo{year}{2016}\natexlab{}.
\newblock \showarticletitle{{PHOG:} Probabilistic Model for Code}. In
  \bibinfo{booktitle}{{\em Proceedings of the 33nd International Conference on
  Machine Learning, {ICML} 2016, New York City, NY, USA, June 19-24, 2016}}.
  \bibinfo{pages}{2933--2942}.
\newblock
\showURL{%
\url{http://jmlr.org/proceedings/papers/v48/bielik16.html}}


\bibitem[\protect\citeauthoryear{Butler, Wermelinger, Yu, and Sharp}{Butler
  et~al\mbox{.}}{2009}]%
        {butler2009}
\bibfield{author}{\bibinfo{person}{S. Butler}, \bibinfo{person}{M.
  Wermelinger}, \bibinfo{person}{Y. Yu}, {and} \bibinfo{person}{H. Sharp}.}
  \bibinfo{year}{2009}\natexlab{}.
\newblock \showarticletitle{Relating Identifier Naming Flaws and Code Quality:
  An Empirical Study}. In \bibinfo{booktitle}{{\em 2009 16th Working Conference
  on Reverse Engineering}}. \bibinfo{pages}{31--35}.
\newblock
\showISSN{1095-1350}
\showDOI{%
\url{https://doi.org/10.1109/WCRE.2009.50}}


\bibitem[\protect\citeauthoryear{Collobert and Weston}{Collobert and
  Weston}{2008}]%
        {collobert2008}
\bibfield{author}{\bibinfo{person}{Ronan Collobert} {and}
  \bibinfo{person}{Jason Weston}.} \bibinfo{year}{2008}\natexlab{}.
\newblock \showarticletitle{A Unified Architecture for Natural Language
  Processing: Deep Neural Networks with Multitask Learning}. In
  \bibinfo{booktitle}{{\em Proceedings of the 25th International Conference on
  Machine Learning}} {\em (\bibinfo{series}{ICML '08})}.
  \bibinfo{publisher}{ACM}, \bibinfo{address}{New York, NY, USA},
  \bibinfo{pages}{160--167}.
\newblock
\showISBNx{978-1-60558-205-4}
\showDOI{%
\url{https://doi.org/10.1145/1390156.1390177}}


\bibitem[\protect\citeauthoryear{Collobert, Weston, Bottou, Karlen,
  Kavukcuoglu, and Kuksa}{Collobert et~al\mbox{.}}{2011}]%
        {collobert2011}
\bibfield{author}{\bibinfo{person}{Ronan Collobert}, \bibinfo{person}{Jason
  Weston}, \bibinfo{person}{L{\'e}on Bottou}, \bibinfo{person}{Michael Karlen},
  \bibinfo{person}{Koray Kavukcuoglu}, {and} \bibinfo{person}{Pavel Kuksa}.}
  \bibinfo{year}{2011}\natexlab{}.
\newblock \showarticletitle{Natural language processing (almost) from scratch}.
\newblock \bibinfo{journal}{{\em Journal of Machine Learning Research\/}}
  \bibinfo{volume}{12}, \bibinfo{number}{Aug} (\bibinfo{year}{2011}),
  \bibinfo{pages}{2493--2537}.
\newblock


\bibitem[\protect\citeauthoryear{Desai, Gulwani, Hingorani, Jain, Karkare,
  Marron, R, and Roy}{Desai et~al\mbox{.}}{2016}]%
        {Desai:2016:PSU:2884781.2884786}
\bibfield{author}{\bibinfo{person}{Aditya Desai}, \bibinfo{person}{Sumit
  Gulwani}, \bibinfo{person}{Vineet Hingorani}, \bibinfo{person}{Nidhi Jain},
  \bibinfo{person}{Amey Karkare}, \bibinfo{person}{Mark Marron},
  \bibinfo{person}{Sailesh R}, {and} \bibinfo{person}{Subhajit Roy}.}
  \bibinfo{year}{2016}\natexlab{}.
\newblock \showarticletitle{Program Synthesis Using Natural Language}. In
  \bibinfo{booktitle}{{\em Proceedings of the 38th International Conference on
  Software Engineering}} {\em (\bibinfo{series}{ICSE '16})}.
  \bibinfo{publisher}{ACM}, \bibinfo{address}{New York, NY, USA},
  \bibinfo{pages}{345--356}.
\newblock
\showISBNx{978-1-4503-3900-1}
\showDOI{%
\url{https://doi.org/10.1145/2884781.2884786}}


\bibitem[\protect\citeauthoryear{Gildea and Jurafsky}{Gildea and
  Jurafsky}{2002}]%
        {gildeajurafsky2002}
\bibfield{author}{\bibinfo{person}{Daniel Gildea} {and} \bibinfo{person}{Daniel
  Jurafsky}.} \bibinfo{year}{2002}\natexlab{}.
\newblock \showarticletitle{Automatic Labeling of Semantic Roles}.
\newblock \bibinfo{journal}{{\em Comput. Linguist.\/}} \bibinfo{volume}{28},
  \bibinfo{number}{3} (\bibinfo{date}{Sept.} \bibinfo{year}{2002}),
  \bibinfo{pages}{245--288}.
\newblock
\showISSN{0891-2017}
\showDOI{%
\url{https://doi.org/10.1162/089120102760275983}}


\bibitem[\protect\citeauthoryear{Grover and Leskovec}{Grover and
  Leskovec}{2016}]%
        {node2vec2016}
\bibfield{author}{\bibinfo{person}{Aditya Grover} {and} \bibinfo{person}{Jure
  Leskovec}.} \bibinfo{year}{2016}\natexlab{}.
\newblock \showarticletitle{node2vec: Scalable Feature Learning for Networks}.
  In \bibinfo{booktitle}{{\em Proceedings of the 22nd ACM SIGKDD International
  Conference on Knowledge Discovery and Data Mining}}.
\newblock


\bibitem[\protect\citeauthoryear{Hastie, Tibshirani, and Friedman}{Hastie
  et~al\mbox{.}}{2001}]%
        {statisticalLearningBook}
\bibfield{author}{\bibinfo{person}{Trevor Hastie}, \bibinfo{person}{Robert
  Tibshirani}, {and} \bibinfo{person}{Jerome Friedman}.}
  \bibinfo{year}{2001}\natexlab{}.
\newblock \bibinfo{booktitle}{{\em The Elements of Statistical Learning}}.
\newblock \bibinfo{publisher}{Springer New York Inc.}, \bibinfo{address}{New
  York, NY, USA}.
\newblock


\bibitem[\protect\citeauthoryear{Hindle, Barr, Su, Gabel, and Devanbu}{Hindle
  et~al\mbox{.}}{2012}]%
        {Hindle:2012:NS:2337223.2337322}
\bibfield{author}{\bibinfo{person}{Abram Hindle}, \bibinfo{person}{Earl~T.
  Barr}, \bibinfo{person}{Zhendong Su}, \bibinfo{person}{Mark Gabel}, {and}
  \bibinfo{person}{Premkumar Devanbu}.} \bibinfo{year}{2012}\natexlab{}.
\newblock \showarticletitle{On the Naturalness of Software}. In
  \bibinfo{booktitle}{{\em Proceedings of the 34th International Conference on
  Software Engineering}} {\em (\bibinfo{series}{ICSE '12})}.
  \bibinfo{publisher}{IEEE Press}, \bibinfo{address}{Piscataway, NJ, USA},
  \bibinfo{pages}{837--847}.
\newblock
\showISBNx{978-1-4673-1067-3}
\showURL{%
\url{http://dl.acm.org/citation.cfm?id=2337223.2337322}}


\bibitem[\protect\citeauthoryear{H{\o}st and {\O}stvold}{H{\o}st and
  {\O}stvold}{2009}]%
        {Host:2009:DMN:1615184.1615204}
\bibfield{author}{\bibinfo{person}{Einar~W. H{\o}st} {and}
  \bibinfo{person}{Bjarte~M. {\O}stvold}.} \bibinfo{year}{2009}\natexlab{}.
\newblock \showarticletitle{Debugging Method Names}. In
  \bibinfo{booktitle}{{\em Proceedings of the 23rd European Conference on ECOOP
  2009 --- Object-Oriented Programming}} {\em (\bibinfo{series}{Genoa})}.
  \bibinfo{publisher}{Springer-Verlag}, \bibinfo{address}{Berlin, Heidelberg},
  \bibinfo{pages}{294--317}.
\newblock
\showISBNx{978-3-642-03012-3}
\showDOI{%
\url{https://doi.org/10.1007/978-3-642-03013-0_14}}


\bibitem[\protect\citeauthoryear{Iyer, Konstas, Cheung, and Zettlemoyer}{Iyer
  et~al\mbox{.}}{2016}]%
        {codenn16}
\bibfield{author}{\bibinfo{person}{Srinivasan Iyer}, \bibinfo{person}{Ioannis
  Konstas}, \bibinfo{person}{Alvin Cheung}, {and} \bibinfo{person}{Luke
  Zettlemoyer}.} \bibinfo{year}{2016}\natexlab{}.
\newblock \showarticletitle{Summarizing Source Code using a Neural Attention
  Model}. In \bibinfo{booktitle}{{\em Proceedings of the 54th Annual Meeting of
  the Association for Computational Linguistics, {ACL} 2016, August 7-12, 2016,
  Berlin, Germany, Volume 1: Long Papers}}.
\newblock
\showURL{%
\url{http://aclweb.org/anthology/P/P16/P16-1195.pdf}}


\bibitem[\protect\citeauthoryear{Karaivanov, Raychev, and Vechev}{Karaivanov
  et~al\mbox{.}}{2014}]%
        {Karaivanov:2014:PST:2661136.2661148}
\bibfield{author}{\bibinfo{person}{Svetoslav Karaivanov},
  \bibinfo{person}{Veselin Raychev}, {and} \bibinfo{person}{Martin Vechev}.}
  \bibinfo{year}{2014}\natexlab{}.
\newblock \showarticletitle{Phrase-Based Statistical Translation of Programming
  Languages}. In \bibinfo{booktitle}{{\em Proceedings of the 2014 ACM
  International Symposium on New Ideas, New Paradigms, and Reflections on
  Programming \& Software}} {\em (\bibinfo{series}{Onward! 2014})}.
  \bibinfo{publisher}{ACM}, \bibinfo{address}{New York, NY, USA},
  \bibinfo{pages}{173--184}.
\newblock
\showISBNx{978-1-4503-3210-1}
\showDOI{%
\url{https://doi.org/10.1145/2661136.2661148}}


\bibitem[\protect\citeauthoryear{Koller, Friedman, Getoor, and Taskar}{Koller
  et~al\mbox{.}}{2007}]%
        {koller2007}
\bibfield{author}{\bibinfo{person}{D. Koller}, \bibinfo{person}{N. Friedman},
  \bibinfo{person}{L. Getoor}, {and} \bibinfo{person}{B. Taskar}.}
  \bibinfo{year}{2007}\natexlab{}.
\newblock \showarticletitle{Graphical Models in a Nutshell}.
\newblock In \bibinfo{booktitle}{{\em Introduction to Statistical Relational
  Learning}}, \bibfield{editor}{\bibinfo{person}{L.~Getoor} {and}
  \bibinfo{person}{B.~Taskar}} (Eds.). \bibinfo{publisher}{MIT Press}.
\newblock


\bibitem[\protect\citeauthoryear{Lafferty, McCallum, and Pereira}{Lafferty
  et~al\mbox{.}}{2001}]%
        {lafferty2001}
\bibfield{author}{\bibinfo{person}{John~D. Lafferty}, \bibinfo{person}{Andrew
  McCallum}, {and} \bibinfo{person}{Fernando C.~N. Pereira}.}
  \bibinfo{year}{2001}\natexlab{}.
\newblock \showarticletitle{Conditional Random Fields: Probabilistic Models for
  Segmenting and Labeling Sequence Data}. In \bibinfo{booktitle}{{\em
  Proceedings of the Eighteenth International Conference on Machine Learning}}
  {\em (\bibinfo{series}{ICML '01})}. \bibinfo{publisher}{Morgan Kaufmann
  Publishers Inc.}, \bibinfo{address}{San Francisco, CA, USA},
  \bibinfo{pages}{282--289}.
\newblock
\showISBNx{1-55860-778-1}
\showURL{%
\url{http://dl.acm.org/citation.cfm?id=645530.655813}}


\bibitem[\protect\citeauthoryear{Levy and Goldberg}{Levy and Goldberg}{2014a}]%
        {levy2014dependency}
\bibfield{author}{\bibinfo{person}{Omer Levy} {and} \bibinfo{person}{Yoav
  Goldberg}.} \bibinfo{year}{2014}\natexlab{a}.
\newblock \showarticletitle{Dependency-Based Word Embeddings.}. In
  \bibinfo{booktitle}{{\em ACL (2)}}. Citeseer, \bibinfo{pages}{302--308}.
\newblock


\bibitem[\protect\citeauthoryear{Levy and Goldberg}{Levy and Goldberg}{2014b}]%
        {levy2014neural}
\bibfield{author}{\bibinfo{person}{Omer Levy} {and} \bibinfo{person}{Yoav
  Goldberg}.} \bibinfo{year}{2014}\natexlab{b}.
\newblock \showarticletitle{Neural Word Embeddings as Implicit Matrix
  Factorization}. In \bibinfo{booktitle}{{\em Advances in Neural Information
  Processing Systems 27: Annual Conference on Neural Information Processing
  Systems 2014, December 8-13 2014, Montreal, Quebec, Canada}}.
  \bibinfo{pages}{2177--2185}.
\newblock


\bibitem[\protect\citeauthoryear{Liblit, Begel, and Sweeser}{Liblit
  et~al\mbox{.}}{2006}]%
        {liblit2006}
\bibfield{author}{\bibinfo{person}{Ben Liblit}, \bibinfo{person}{Andrew Begel},
  {and} \bibinfo{person}{Eve Sweeser}.} \bibinfo{year}{2006}\natexlab{}.
\newblock \showarticletitle{Cognitive Perspectives on the Role of Naming in
  Computer Programs}. In \bibinfo{booktitle}{{\em Proceedings of the 18th
  Annual Psychology of Programming Workshop}}. Psychology of Programming
  Interest Group, \bibinfo{address}{Sussex, United Kingdom}.
\newblock


\bibitem[\protect\citeauthoryear{Lopes, Maj, Martins, Saini, Yang, Zitny,
  Sajnani, and Vitek}{Lopes et~al\mbox{.}}{2017}]%
        {dejavu2017}
\bibfield{author}{\bibinfo{person}{Cristina~V. Lopes}, \bibinfo{person}{Petr
  Maj}, \bibinfo{person}{Pedro Martins}, \bibinfo{person}{Vaibhav Saini},
  \bibinfo{person}{Di Yang}, \bibinfo{person}{Jakub Zitny},
  \bibinfo{person}{Hitesh Sajnani}, {and} \bibinfo{person}{Jan Vitek}.}
  \bibinfo{year}{2017}\natexlab{}.
\newblock \showarticletitle{D{\'e}J\`{a}Vu: A Map of Code Duplicates on
  GitHub}.
\newblock \bibinfo{journal}{{\em Proc. ACM Program. Lang.\/}}
  \bibinfo{volume}{1}, \bibinfo{number}{OOPSLA}, Article
  \bibinfo{articleno}{84} (\bibinfo{date}{Oct.} \bibinfo{year}{2017}),
  \bibinfo{numpages}{28}~pages.
\newblock
\showISSN{2475-1421}
\showDOI{%
\url{https://doi.org/10.1145/3133908}}


\bibitem[\protect\citeauthoryear{Maddison and Tarlow}{Maddison and
  Tarlow}{2014}]%
        {Maddison:2014:SGM:3044805.3044965}
\bibfield{author}{\bibinfo{person}{Chris~J. Maddison} {and}
  \bibinfo{person}{Daniel Tarlow}.} \bibinfo{year}{2014}\natexlab{}.
\newblock \showarticletitle{Structured Generative Models of Natural Source
  Code}. In \bibinfo{booktitle}{{\em Proceedings of the 31st International
  Conference on International Conference on Machine Learning - Volume 32}} {\em
  (\bibinfo{series}{ICML'14})}. \bibinfo{publisher}{JMLR.org},
  \bibinfo{pages}{II--649--II--657}.
\newblock
\showURL{%
\url{http://dl.acm.org/citation.cfm?id=3044805.3044965}}


\bibitem[\protect\citeauthoryear{Melamud, Levy, and Dagan}{Melamud
  et~al\mbox{.}}{2015}]%
        {melamud15}
\bibfield{author}{\bibinfo{person}{Oren Melamud}, \bibinfo{person}{Omer Levy},
  {and} \bibinfo{person}{Ido Dagan}.} \bibinfo{year}{2015}\natexlab{}.
\newblock \showarticletitle{A Simple Word Embedding Model for Lexical
  Substitution}. In \bibinfo{booktitle}{{\em Proceedings of the 1st Workshop on
  Vector Space Modeling for Natural Language Processing, VS@NAACL-HLT 2015,
  June 5, 2015, Denver, Colorado, {USA}}}. \bibinfo{pages}{1--7}.
\newblock
\showURL{%
\url{http://aclweb.org/anthology/W/W15/W15-1501.pdf}}


\bibitem[\protect\citeauthoryear{Mikolov, Chen, Corrado, and Dean}{Mikolov
  et~al\mbox{.}}{2013a}]%
        {mikolovEfficient2013}
\bibfield{author}{\bibinfo{person}{Tomas Mikolov}, \bibinfo{person}{Kai Chen},
  \bibinfo{person}{Greg Corrado}, {and} \bibinfo{person}{Jeffrey Dean}.}
  \bibinfo{year}{2013}\natexlab{a}.
\newblock \showarticletitle{Efficient Estimation of Word Representations in
  Vector Space}.
\newblock \bibinfo{journal}{{\em CoRR\/}}  \bibinfo{volume}{abs/1301.3781}
  (\bibinfo{year}{2013}).
\newblock
\showURL{%
\url{http://arxiv.org/abs/1301.3781}}


\bibitem[\protect\citeauthoryear{Mikolov, Sutskever, Chen, Corrado, and
  Dean}{Mikolov et~al\mbox{.}}{2013b}]%
        {mikolovDistributed2013}
\bibfield{author}{\bibinfo{person}{Tomas Mikolov}, \bibinfo{person}{Ilya
  Sutskever}, \bibinfo{person}{Kai Chen}, \bibinfo{person}{Greg Corrado}, {and}
  \bibinfo{person}{Jeffrey Dean}.} \bibinfo{year}{2013}\natexlab{b}.
\newblock \showarticletitle{Distributed Representations of Words and Phrases
  and Their Compositionality}. In \bibinfo{booktitle}{{\em Proceedings of the
  26th International Conference on Neural Information Processing Systems}} {\em
  (\bibinfo{series}{NIPS'13})}. \bibinfo{publisher}{Curran Associates Inc.},
  \bibinfo{address}{USA}, \bibinfo{pages}{3111--3119}.
\newblock
\showURL{%
\url{http://dl.acm.org/citation.cfm?id=2999792.2999959}}


\bibitem[\protect\citeauthoryear{Mou, Li, Zhang, Wang, and Jin}{Mou
  et~al\mbox{.}}{2016}]%
        {tbcnn2016}
\bibfield{author}{\bibinfo{person}{Lili Mou}, \bibinfo{person}{Ge Li},
  \bibinfo{person}{Lu Zhang}, \bibinfo{person}{Tao Wang}, {and}
  \bibinfo{person}{Zhi Jin}.} \bibinfo{year}{2016}\natexlab{}.
\newblock \showarticletitle{Convolutional Neural Networks over Tree Structures
  for Programming Language Processing}. In \bibinfo{booktitle}{{\em Proceedings
  of the Thirtieth AAAI Conference on Artificial Intelligence}} {\em
  (\bibinfo{series}{AAAI'16})}. \bibinfo{publisher}{AAAI Press},
  \bibinfo{pages}{1287--1293}.
\newblock
\showURL{%
\url{http://dl.acm.org/citation.cfm?id=3015812.3016002}}


\bibitem[\protect\citeauthoryear{Pearl}{Pearl}{2011}]%
        {pearl2011bayesian}
\bibfield{author}{\bibinfo{person}{Judea Pearl}.}
  \bibinfo{year}{2011}\natexlab{}.
\newblock \showarticletitle{Bayesian networks}.
\newblock  (\bibinfo{year}{2011}).
\newblock


\bibitem[\protect\citeauthoryear{Pearl}{Pearl}{2014}]%
        {pearl2014probabilistic}
\bibfield{author}{\bibinfo{person}{Judea Pearl}.}
  \bibinfo{year}{2014}\natexlab{}.
\newblock \bibinfo{booktitle}{{\em Probabilistic reasoning in intelligent
  systems: networks of plausible inference}}.
\newblock \bibinfo{publisher}{Elsevier}.
\newblock


\bibitem[\protect\citeauthoryear{Pennington, Socher, and Manning}{Pennington
  et~al\mbox{.}}{2014}]%
        {glove2014}
\bibfield{author}{\bibinfo{person}{Jeffrey Pennington},
  \bibinfo{person}{Richard Socher}, {and} \bibinfo{person}{Christopher~D.
  Manning}.} \bibinfo{year}{2014}\natexlab{}.
\newblock \showarticletitle{GloVe: Global Vectors for Word Representation}. In
  \bibinfo{booktitle}{{\em Empirical Methods in Natural Language Processing
  (EMNLP)}}. \bibinfo{pages}{1532--1543}.
\newblock
\showURL{%
\url{http://www.aclweb.org/anthology/D14-1162}}


\bibitem[\protect\citeauthoryear{Raychev, Bielik, and Vechev}{Raychev
  et~al\mbox{.}}{2016a}]%
        {decisionTrees2016}
\bibfield{author}{\bibinfo{person}{Veselin Raychev}, \bibinfo{person}{Pavol
  Bielik}, {and} \bibinfo{person}{Martin Vechev}.}
  \bibinfo{year}{2016}\natexlab{a}.
\newblock \showarticletitle{Probabilistic Model for Code with Decision Trees}.
  In \bibinfo{booktitle}{{\em Proceedings of the 2016 ACM SIGPLAN International
  Conference on Object-Oriented Programming, Systems, Languages, and
  Applications}} {\em (\bibinfo{series}{OOPSLA 2016})}.
  \bibinfo{publisher}{ACM}, \bibinfo{address}{New York, NY, USA},
  \bibinfo{pages}{731--747}.
\newblock
\showISBNx{978-1-4503-4444-9}
\showDOI{%
\url{https://doi.org/10.1145/2983990.2984041}}


\bibitem[\protect\citeauthoryear{Raychev, Bielik, Vechev, and Krause}{Raychev
  et~al\mbox{.}}{2016b}]%
        {raychev2016noisy}
\bibfield{author}{\bibinfo{person}{Veselin Raychev}, \bibinfo{person}{Pavol
  Bielik}, \bibinfo{person}{Martin Vechev}, {and} \bibinfo{person}{Andreas
  Krause}.} \bibinfo{year}{2016}\natexlab{b}.
\newblock \showarticletitle{Learning Programs from Noisy Data}. In
  \bibinfo{booktitle}{{\em Proceedings of the 43rd Annual ACM SIGPLAN-SIGACT
  Symposium on Principles of Programming Languages}} {\em
  (\bibinfo{series}{POPL '16})}. \bibinfo{publisher}{ACM},
  \bibinfo{address}{New York, NY, USA}, \bibinfo{pages}{761--774}.
\newblock
\showISBNx{978-1-4503-3549-2}
\showDOI{%
\url{https://doi.org/10.1145/2837614.2837671}}


\bibitem[\protect\citeauthoryear{Raychev, Vechev, and Krause}{Raychev
  et~al\mbox{.}}{2015}]%
        {jsnice2015}
\bibfield{author}{\bibinfo{person}{Veselin Raychev}, \bibinfo{person}{Martin
  Vechev}, {and} \bibinfo{person}{Andreas Krause}.}
  \bibinfo{year}{2015}\natexlab{}.
\newblock \showarticletitle{Predicting Program Properties from "Big Code"}. In
  \bibinfo{booktitle}{{\em Proceedings of the 42Nd Annual ACM SIGPLAN-SIGACT
  Symposium on Principles of Programming Languages}} {\em
  (\bibinfo{series}{POPL '15})}. \bibinfo{publisher}{ACM},
  \bibinfo{address}{New York, NY, USA}, \bibinfo{pages}{111--124}.
\newblock
\showISBNx{978-1-4503-3300-9}
\showDOI{%
\url{https://doi.org/10.1145/2676726.2677009}}


\bibitem[\protect\citeauthoryear{Raychev, Vechev, and Yahav}{Raychev
  et~al\mbox{.}}{2014}]%
        {raychev14}
\bibfield{author}{\bibinfo{person}{Veselin Raychev}, \bibinfo{person}{Martin
  Vechev}, {and} \bibinfo{person}{Eran Yahav}.}
  \bibinfo{year}{2014}\natexlab{}.
\newblock \showarticletitle{Code Completion with Statistical Language Models}.
\newblock \bibinfo{journal}{{\em SIGPLAN Not.\/}} \bibinfo{volume}{49},
  \bibinfo{number}{6} (\bibinfo{date}{June} \bibinfo{year}{2014}),
  \bibinfo{pages}{419--428}.
\newblock
\showISSN{0362-1340}
\showDOI{%
\url{https://doi.org/10.1145/2666356.2594321}}


\bibitem[\protect\citeauthoryear{Shalev-Shwartz and Ben-David}{Shalev-Shwartz
  and Ben-David}{2014}]%
        {shwartz2014}
\bibfield{author}{\bibinfo{person}{Shai Shalev-Shwartz} {and}
  \bibinfo{person}{Shai Ben-David}.} \bibinfo{year}{2014}\natexlab{}.
\newblock \bibinfo{booktitle}{{\em Understanding Machine Learning: From Theory
  to Algorithms}}.
\newblock \bibinfo{publisher}{Cambridge University Press},
  \bibinfo{address}{New York, NY, USA}.
\newblock
\showISBNx{1107057132, 9781107057135}


\bibitem[\protect\citeauthoryear{Sutton and McCallum}{Sutton and
  McCallum}{2012}]%
        {sutton2012}
\bibfield{author}{\bibinfo{person}{Charles Sutton} {and}
  \bibinfo{person}{Andrew McCallum}.} \bibinfo{year}{2012}\natexlab{}.
\newblock \showarticletitle{An Introduction to Conditional Random Fields}.
\newblock \bibinfo{journal}{{\em Found. Trends Mach. Learn.\/}}
  \bibinfo{volume}{4}, \bibinfo{number}{4} (\bibinfo{date}{April}
  \bibinfo{year}{2012}), \bibinfo{pages}{267--373}.
\newblock
\showISSN{1935-8237}
\showDOI{%
\url{https://doi.org/10.1561/2200000013}}


\bibitem[\protect\citeauthoryear{Takang, Grubb, and Macredie}{Takang
  et~al\mbox{.}}{1996}]%
        {takang96}
\bibfield{author}{\bibinfo{person}{Armstrong~A. Takang},
  \bibinfo{person}{Penny~A. Grubb}, {and} \bibinfo{person}{Robert~D.
  Macredie}.} \bibinfo{year}{1996}\natexlab{}.
\newblock \showarticletitle{The effects of comments and identifier names on
  program comprehensibility: an experimental investigation}.
\newblock \bibinfo{journal}{{\em J. Prog. Lang.\/}} \bibinfo{volume}{4},
  \bibinfo{number}{3} (\bibinfo{year}{1996}), \bibinfo{pages}{143--167}.
\newblock
\showURL{%
\url{http://compscinet.dcs.kcl.ac.uk/JP/jp040302.abs.html}}


\bibitem[\protect\citeauthoryear{Turian, Ratinov, and Bengio}{Turian
  et~al\mbox{.}}{2010}]%
        {turian2010}
\bibfield{author}{\bibinfo{person}{Joseph Turian}, \bibinfo{person}{Lev
  Ratinov}, {and} \bibinfo{person}{Yoshua Bengio}.}
  \bibinfo{year}{2010}\natexlab{}.
\newblock \showarticletitle{Word Representations: A Simple and General Method
  for Semi-supervised Learning}. In \bibinfo{booktitle}{{\em Proceedings of the
  48th Annual Meeting of the Association for Computational Linguistics}} {\em
  (\bibinfo{series}{ACL '10})}. \bibinfo{publisher}{Association for
  Computational Linguistics}, \bibinfo{address}{Stroudsburg, PA, USA},
  \bibinfo{pages}{384--394}.
\newblock
\showURL{%
\url{http://dl.acm.org/citation.cfm?id=1858681.1858721}}


\bibitem[\protect\citeauthoryear{Zilberstein and Yahav}{Zilberstein and
  Yahav}{2016}]%
        {Zilberstein:2016:LCN:2986012.2986013}
\bibfield{author}{\bibinfo{person}{Meital Zilberstein} {and}
  \bibinfo{person}{Eran Yahav}.} \bibinfo{year}{2016}\natexlab{}.
\newblock \showarticletitle{Leveraging a Corpus of Natural Language
  Descriptions for Program Similarity}. In \bibinfo{booktitle}{{\em Proceedings
  of the 2016 ACM International Symposium on New Ideas, New Paradigms, and
  Reflections on Programming and Software}} {\em (\bibinfo{series}{Onward!
  2016})}. \bibinfo{publisher}{ACM}, \bibinfo{address}{New York, NY, USA},
  \bibinfo{pages}{197--211}.
\newblock
\showISBNx{978-1-4503-4076-2}
\showDOI{%
\url{https://doi.org/10.1145/2986012.2986013}}


\end{thebibliography}
